 \documentclass[twoside,reqno,11pt]{fcaa-var} 

\usepackage{graphicx}
\usepackage{epsfig}

\usepackage{amsthm}
\usepackage{amsmath}
\usepackage{latexsym}
\usepackage{amsfonts}
\usepackage{amssymb}

\newtheoremstyle{theorem}
  {15pt}          
  {15pt}  
  {\sl}  
  {\parindent}
  {\sc}  
  {. }   
  { }    
  {}     
\theoremstyle{theorem}

\newtheoremstyle{defi}
  {15pt}          
  {15pt}  
  {\rm}  
  {\parindent}     
  {\sc}  
  {. }    
  { }    
  {}     
\theoremstyle{defi}

 \def\proofname{\indent\rm P\ r\ o\ o\ f.}
 
 \def\qed{\hfill$\Box$} 

 \usepackage{hyperref} 
 \def\theequation{\arabic{section}.\arabic{equation}}

\usepackage{graphicx}
\usepackage{epsfig}
\usepackage{amsmath}
\usepackage{amssymb}
\usepackage{mathtools}

\DeclarePairedDelimiter{\ceil}{\lceil}{\rceil}

\newcommand{\ds}{\displaystyle }

 \def\themycopyrightfootnote{\vspace*{3pt}
   \par  \noindent {\it Fract. Calc. Appl. Anal.}, Vol. {\bf 23}, No 3 (2020), pp. 656–693;\\
{\bf doi:} 10.1515/fca-2020-0034.
   \hfill  \vspace*{-36pt}
  }

 \title[Generalized fractional Poisson process and \dots]
       {Generalized fractional Poisson process and \\ [3pt] related stochastic dynamics}

 \author[\normalsize T.M. Michelitsch, A.P. Riascos]
        {\normalsize Thomas M. Michelitsch $^{1, \S}$,  Alejandro P. Riascos $^2$}

\begin{document}

 \vbox to 1.5cm { \vfill }


 \bigskip \medskip

 \begin{abstract}
 
We survey the `generalized fractional Poisson process' (GFPP). The GFPP is a renewal process generalizing Laskin's fractional Poisson counting process and was first introduced by Cahoy and Polito.
The GFPP contains two index parameters with admissible ranges $0<\beta\leq 1$, $\alpha >0$ and a parameter characterizing the time scale. The GFPP involves Prabhakar 
generalized Mittag-Leffler functions and contains for special choices of the parameters
the Laskin fractional Poisson process, the Erlang process and the standard Poisson process. 
We demonstrate this by means of explicit formulas.
We develop the Montroll-Weiss continuous-time random walk (CTRW) for the GFPP on undirected networks which has Prabhakar distributed waiting times between the jumps of the walker.
For this walk, we derive a generalized fractional Kolmogorov-Feller equation which involves Prabhakar generalized fractional operators governing the stochastic motions on the network.
We analyze in $d$ dimensions the `well-scaled' diffusion limit and obtain a fractional diffusion equation which is of the same type as for a walk with Mittag-Leffler distributed waiting times.
The GFPP has the potential to capture various aspects in the dynamics of certain complex systems.

 \medskip


 \smallskip


 \end{abstract}

 \maketitle

 \vspace*{-25pt}


\section{Introduction} 

\setcounter{section}{1}
\setcounter{equation}{0}\setcounter{theorem}{0}

A characteristic feature in
many complex systems are non-exponential time patterns with asymptotic power-law behavior as occurring
for instance in anomalous
transport and diffusion phenomena \cite{Gorenflo2010,ZumofenShlesingerKlafter1996,MetzlerKlafter2000,SaichevZaslavski1997,Shlesinger2017,Zaslavsky2002} (and many others).
On the other hand, it has been found that random walk approaches have the capacity to capture various dynamic features
of certain complex systems. Markovian random walk models have been established to investigate a variety of aspects of stochastic processes
in continuous spaces
and networks as various as
human mobility patterns \cite{RiascosMateos2017} and random motions with long-range jumps on undirected networks
with the emergence of L\'evy flights
\cite{TMM-APR-ISTE2019,TMM-APR-JPhys-A2017,RiascosMateos2015,RiascosMichel2018}.

However, Markovian approaches are not able to capture (long-time) memory features.
For the description of the stochastic behavior of such systems non-Markovian models
of time fractional diffusion and fractional
Fokker Planck equations have been developed exhibiting non-Markovian long-time memory effects \cite{Gorenflo2007,MetzlerKlafter2004,MetzlerKlafter2000,Zaslavsky2002}.

Continuous-time random walks (CTRWs) are stochastic processes where the walker makes jumps at random times which follow a renewal process.
The CTRW approach was introduced by Montroll and Weiss in their 1965
seminal paper \cite{MontrollWeiss1965}.
The notion of {\it Continuous Time Random Walk - CTRW }
was coined by Scher and Lax \cite{ScherLax1973} who launched in this way the field of `anomalous transport' \cite{Shlesinger2017}.
The CTRW mathematically is equivalent to a
compound renewal process or a random walk subordinated to a renewal process  \cite{Cox1967,Gorenflo2010}.
The CTRW meanwhile has been applied to a large variety of problems including charge transport in disordered systems
\cite{ScherMontroll1975}, the stochastic processes in protein folding \cite{Brungelson1989},
the dynamics of chemical reactions \cite{SungBarkaiSilbey2002}, glass phenomenology \cite{MonthusBouchaud1996}, transport phenomena in low dimensional
chaotic systems \cite{ZumofenShlesingerKlafter1996,SolomonSwinner1993}
and generally in anomalous transport models \cite{Angstmann-et-al2017,KlagesRadonsSokolov2008}.

On the other hand, a huge literature exists \cite{MetzlerKlafter2000} with various generalizations of the CTRW approach
including variants leading to
a fractional Fokker Planck equation \cite{BarkaiMetzlerKlafter2000} and
the so-called `Aging Continuous Time Random Walk' represents a variant where the stochastic properties of the walk
depend on the `age' of the walk \cite{BarkaiCheng2003}.
The CTRW approach also was generalized to the case that waiting times of successive jumps are correlated by memory kernels \cite{ChechkinHofmannSokolov2009}.

The underlying renewal process in classical CTRWs is the Poisson process with exponential waiting time distributions
and the Markov property \cite{Feller1971,MetzlerKlafter2000}. However, as mentioned it has turned out that many `complex systems' exhibit time patterns with
heavy-tailed power-law behavior rather than exponential waiting time characteristics \cite{Laskin2003,MetzlerKlafter2000}.

A non-Markovian fractional generalization of the Markovian Poisson process was introduced and analyzed by Laskin \cite{Laskin2003}
who called this process `{\it fractional Poisson process}'. He demonstrated the utmost importance of the fractional Poisson distribution in several applications \cite{Laskin2009}. Various properties of the fractional Poisson process
were meanwhile analyzed by several authors
\cite{BeghinOrsinger2009,GorenfloMainardi2013,HilferAnton1995,MainardiGorenfloScalas2004,MeerschaertEtal2011,RepinSaichev2000}  and consult also \cite{OrsingherPolito2011,PolitoScalas2016} where space-time variants of the fractional Poisson process are introduced.
The Laskin fractional Poisson process is a renewal process with Mittag-Leffler distributed waiting times exhibiting
asymptotic heavy-tailed power law arrival densities with long-time memory effects as observed in
a wide class of complex systems \cite{MetzlerKlafter2000,SaichevZaslavski1997,Zaslavsky2002}.

The heavy-tailed power-law asymptotic features of the Mittag-Leffler waiting time law
in CTRWs and asymptotic universality of Mittag-Leffler pattern for large observation times
among other properties were thoroughly analyzed in the references \cite{BeghinOrsinger2009,GorenfloKilbas2014,Gorenflo2010,GorenfloMainardi2006}.

The subject of the present survey paper is a generalization of Laskin's fractional Poisson counting process and the analysis of the resulting stochastic dynamics on graphs.
This generalization first was proposed by Cahoy and Polito in a seminal paper \cite{CahoyPolito2013}.
We call this process the {\it `generalized fractional Poisson process' (GFPP)}. Both the GFPP like the fractional Poisson process exhibit non-Markovian and long-memory effects.

The paper is organized as follows. In the first part (Section \ref{CTRW})
we recall general concepts of renewal theory employed in the Continuous Time Random Walk (CTRW)
of Montroll and Weiss \cite{MontrollWeiss1965}.
Based on this approach we recall in Section \ref{FPP}
the Laskin fractional Poisson process and distribution \cite{Laskin2003,SaichevZaslavski1997}.
Section \ref{generalized} is devoted to the
GFPP counting process. We derive its waiting time distribution with related quantities
and analyze their asymptotic behaviors.
In Section \ref{GeneralizedFractPoissonDistri} the probability distribution
for $n$ arrivals within a time interval (`state-probabilities')
is deduced for the GFPP. We refer this distribution to as the {\it `generalized fractional Poisson distribution' (GFPD)}.
We derive in Section \ref{GenFractionalPoissonProcess} the expected number of arrivals in a GFPP and analyze asymptotic power-law features.
The GFPP contains two index parameters with admissible ranges $0<\beta\leq 1$ and $\alpha>0$ and a parameter
characterizing the time scale.
The Laskin fractional Poisson process and the
standard Poisson process are contained as special cases in the GFPP.

As an application we consider a CTRW on undirected networks (Section \ref{CTRWnetworks}). In this part we subordinate a random walk on the network to the GFPP:
We derive (Section \ref{definitionGFPP}) the `generalized fractional Kolmogorov-Feller equation'
that governs the resulting stochastic motion. We show that for $\alpha=1$ the Laskin fractional Kolmogorov-Feller equations
are recovered as well as for $\alpha=1$, $\beta=1$ the classical Kolmogorov-Feller equations of the standard Poisson process.
Finally in Section \ref{genfractdiffusion} we analyze stochastic motions subordinated to a GFPP
in the infinite $d$-dimensional integer lattice and derive in a `well-scaled diffusion limit' the governing diffusion equation which is of the
same fractional type as for the fractional Poisson process. For $\alpha=1$, $\beta=1$ this equation reduces to the standard diffusion equation (Fick's second law) of normal diffusion.

The entire demonstration is accompanied by supplementary materials/Appendices
where detailed mathematical derivations are performed and properties are discussed.
\vspace*{-4pt} 
\section{Renewal processes and waiting time distributions} \label{CTRW}

\setcounter{section}{2}
\setcounter{equation}{0}\setcounter{theorem}{0} 

\subsection{Preliminaries} 

In order to analyze the GFPP counting process
let us recall some basic properties of
renewal processes and of the Montroll-Weiss continuous-time random walk (CTRW) approach.
In the present paper, we deal with {\it causal} generalized functions and distributions \cite{GelfangShilov1968}.
A function or distribution is referred
to as {\it causal} if it has the representation $\Theta(t)f(t)$ where $\Theta(t)$
is the Heaviside step function $\Theta(t)=1$ for $t\geq 0$ and $\Theta(t)=0$ for $t<0$.
Throughout the paper we utilize Laplace transforms of causal functions and distributions 
which we define by Eq. (\ref{laplacetranfocausalfunction}) in Appendix \ref{AppendLaplacetrafo} where
we briefly outline the properties relevant in our analysis. We employ notation $\mathcal{L}\{f(t)\}=:{\tilde f}(s)$
for the Laplace transform of a causal function $\Theta(t)f(t)$.
\smallskip 

Let us now evoke the basic features of
renewal processes and waiting time distributions as a basis of the Montroll-Weiss continuous time random walk (CTRW)
approach \cite{MontrollWeiss1965,ScherLax1973} and consult also \cite{GorenfloMainardi2013,Gorenflo2010,GorenfloMainardiVivoli2007,MainardiGorenfloScalas2004}.
We assume that a walker makes jumps at random times $0 < t_1 < t_2 < \ldots t_n < \ldots \infty$ on a set of accessible sites. The walk can be conceived as a stream of jump events that happen at `arrival times' $t_k$.
We start the observation of the walk at time $t=0$ where the walker is sitting on his initial
position until making the first jump at $t=t_1$ (arrival of the first jump event) and so forth.
The waiting times $\Delta t_k = t_k-t_{k-1} > 0$ between two successive (jump-) events are assumed to be drawn from the same probability density distribution $\chi(t)$.
The waiting times $\Delta t_k$ then
are independent and identically distributed (IID) random variables.
$\chi(\tau) {\rm d}\tau $ indicates the probability that the first jump event (first arrival) happens exactly at time $\tau$.
We refer $\chi(t)$ to as {\it waiting time probability density function} (waiting time PDF)
or short `{\it jump density}'.
The probability that the walker has jumped at least once within the time interval $[0,t]$ is given by
the cumulative probability distribution
\vskip -13pt%
\begin{equation}
 \label{waitintimedis}
prob(\Delta t \leq t)= \Psi(t) = \int_0^t \chi(\tau){\rm d}\tau ,\hspace{0.5cm} t\geq 0 ,
\hspace{0.5cm} \lim_{t\rightarrow\infty} \Psi(t) =1-0,
\end{equation}
\vskip -3pt \noindent%
with the obvious initial condition $\Psi(t=0)=0$. (\ref{waitintimedis}) is also called `{\it failure probability}'. The probability that no jump event happened within $[0,t]$, i.e. that the waiting time is greater than $t$ is given by
\vskip -10pt %
\begin{equation}
 \label{waiting-timeprob}
 prob(\Delta t > t) = \Phi^{(0}(t)=1-\Psi(t) =\int_t^{\infty}\chi(\tau){\rm d}\tau .
\end{equation}
\vskip -3pt \noindent %
The cumulative probability distribution (\ref{waiting-timeprob}) that the walker during $[0,t]$ still is waiting on
its initial site also is called `{\it survival probability}'. We notice that densities such as the jump density $\chi(t)$
have physical dimension $[sec]^{-1}$,
whereas the cumulative distributions  (\ref{waitintimedis}), (\ref{waiting-timeprob}) are dimensionless probabilities.
The jump density $\chi(\tau)$ is non-negative and normalized (See (\ref{waitintimedis})), i.e. for $t\rightarrow \infty$
the walker (almost) surely has made (at least) one jump. This means that survival probability (\ref{waiting-timeprob})
tends to zero as time approaches infinity $\lim_{t\rightarrow\infty} \Phi^{(0}(t) \rightarrow 0$.

What is the PDF that the walker makes its second jump exactly at time $t$? Since the waiting times
$\Delta t_1$ and $\Delta t_2$ are IID the probability that the walker makes its first jump exactly at $\Delta t_1$
and the second jump at the time $\Delta t_1+\Delta t_2$ is ${\rm d}\chi^{(2)} =\chi(\Delta t_1)\chi(\Delta t_2){\rm d t}_1{\rm d}t_2$.
Integrating this quantity over the line $t=\Delta t_1+\Delta t_2$
(and dividing by a time increment) yields the PDF (probability per unit time)
that the second step is made exactly at time $t$, namely
\vskip - 14pt 
\begin{align}\nonumber
 \chi^{(2)}(t)&= \int_0^t\chi(\tau)\chi(t-\tau){\rm d}\tau\\
 &=
 \int_0^{\infty}\int_0^{\infty}\chi(\tau_1)\chi(\tau_2)\delta\left(t-(\tau_1+\tau_2)\right){\rm d}\tau_1{\rm d}\tau_2 , \label{deltachi2}
\end{align}
\vskip -3pt \noindent %
where $\delta(u)$ indicates Dirac's $\delta$-distribution.
In this relation we used that $\chi(t)=\chi(t)\Theta(t)$ is as a {\it causal function}.
The PDF $\chi^{(n)}(t)$ that the walker makes its
$n$th jump {\it exactly at time $t$} then is
\vskip - 13pt%
\begin{equation}
 \label{probofnjumps}
 \chi^{(n)}(t) = \int_0^{\infty}\ldots \int_0^{\infty}\chi(\tau_1)\ldots \chi(\tau_n)\delta\left(t-\sum_{j=1}^n\tau_j\right){\rm d}\tau_1\ldots {\rm d}\tau_n.
\end{equation}
\vskip -3pt \noindent %
We refer (\ref{probofnjumps}) to as $n$-jump probability density.
The physical dimension of the $n$-step PDF $\chi^{(n)}(t)$ (for all $n$) is $[sec]^{-1}$
whereas the cumulative
probability $\Psi(t)$ of (\ref{waitintimedis}) is dimensionless.
It is important to note that the PDF $\chi^{(n)}(t)$ to make the $n$th jump exactly at time $t$
is not the same as the dimensionless {\it probability} $\Phi^{(n)}(t)$ that the
walker within the time interval $[0,t]$ has made $n$ jumps.
In the latter case the walker may have made the $n$th step at a time $\tau<t$
and waited for $t-\tau$ (with probability $\Phi^{(0)}(t-\tau)=1-\Psi(t-\tau)$).
The probability for $n$ arrivals within the time interval $[0,t]$ is hence given by
\vskip -11pt %
\begin{equation}
 \label{nstepprobabilityuptot}
 \Phi^{(n)}(t) = \int_0^t\left(1-\Psi(t-\tau)\right)\chi^{(n)}(\tau){\rm d}\tau =
 \int_0^t \Phi^{(0)}(t-\tau)\chi^{(n)}(\tau){\rm d}\tau,
\end{equation}
%
for $n=0,1,2,\dots$. The probabilities (\ref{nstepprobabilityuptot}) also are referred to as `state-probabilities'.
This convolution covers also $n=0$ when we account for $\chi^{(0)}(t)=\delta(t)$.
It follows that
$\Phi^{(n)}(t)$ ($n=0,1,2,\ldots $) is a dimensionless probability distribution and normalized
on the `state-space' $n\in \mathbb{N}_0$, namely
\vskip -10pt%
\begin{equation}
 \label{phinnormaliszation}
 \sum_{n=0}^{\infty} \Phi^{(n)}(t) = 1 ,
\end{equation}
\vskip -1pt \noindent %
i.e., at any time $t \geq 0$ the walker for sure either waits, i.e. has performed $n=0$ jumps,
or has performed $n=1,2,\ldots ,\infty$ jumps. The normalization
(\ref{phinnormaliszation}) is proved subsequently.
It is straightforward to see that both (\ref{probofnjumps}) and (\ref{nstepprobabilityuptot}) fulfill the recursion
\vskip -10pt%
\begin{equation}
 \label{nstepssteps}
 \begin{array}{l}
\ds  \chi^{(n)}(t) = \int_0^t \chi^{(n-1)}(\tau)\chi(t-\tau){\rm d}\tau ,\hspace{1cm} \chi^{(0)}(t)=\delta(t), \\ \\
 \ds \Phi^{(n)}(t) = \int_0^t \Phi^{(n-1)}(\tau)\chi(t-\tau){\rm d}\tau  ,\hspace{1cm} \Phi^{(0)}(t)=1-\Psi(t),
 \end{array}
\end{equation}
\vskip -3pt \noindent %
where $n=1$ recovers $\chi_1(t)=\chi(t)=\int_0^t\chi(t-\tau)\delta(\tau){\rm d}\tau$.
In all relations we account for causality $\chi(t)= \Theta(t)\chi(t)$.
The $n$-jump PDF $\chi^{(n)}(t)$ of (\ref{probofnjumps}) has the Laplace transform
\vskip - 12pt%
\begin{align}\nonumber
{\tilde \chi}^{(n)}(s) &= \int_0^{\infty}{\rm d}\tau_1 \ldots \int_0^{\infty}{\rm d}\tau_n  \chi(\tau_1)\ldots \chi(\tau_n)\int_{0}^{\infty}{\rm d} t e^{-st} \delta\left(t-\sum_{j=1}^n\tau_j\right)\\ \nonumber
& = \int_0^{\infty}\ldots \int_0^{\infty}\chi(\tau_1)\ldots \chi(\tau_n) {\rm d}\tau_1\ldots {\rm d}\tau_n e^{-s\sum_{j=1}^n\tau_j} \\
&= \left(\int_0^{\infty}\chi(\tau)e^{-s\tau}{\rm d}\tau\right)^n= ({\tilde \chi}(s))^n,
\label{laplacenjumps}
\end{align}
\vskip -2pt \noindent %
reflecting the product rule for convolutions where ${\tilde \chi}(s)$ indicates the Laplace
transform of the (single-) jump density $\chi(t)=\chi^{(1)}(t)$. The property
${\tilde \chi}^{(n)}(s)|_{s=0} = \int_0^{\infty} \chi^{(n)}(t){\rm d}t =1 $ in (\ref{laplacenjumps})
reflects the normalization condition, i.e. for $t\rightarrow\infty$ the walker has made $n=0,1,2,\ldots ,\infty$ jumps almost surely.
We then get for the Laplace transform of (\ref{nstepprobabilityuptot})
\vskip -10pt%
\begin{equation}
 \label{laplansteps}
 {\tilde \Phi}^{(n)}(s) =\frac{1-{\tilde \chi}(s)}{s}({\tilde \chi}(s))^n =\frac{1}{s}\left({\tilde \chi}^{(n)}(s)-{\tilde \chi}^{(n+1)}(s)\right), \hspace{0.5cm} n \in \mathbb{N}_0 ,
\end{equation}
\vskip -2pt \noindent %
where for $n=0$ we have for the Laplace transform of the survival probability ${\tilde \Phi}^{(0)}(s) = \frac{1-{\tilde \chi}(s)}{s}$ when
we account for ${\chi}_0(t)=\delta(t)$ with ${\tilde \chi}^{(0)}(s) =\mathcal{L}\left\{\delta(t)\right\}=1$.
It is important to notice that
$|{\tilde \chi}(s)| \leq {\chi (s=0)}= 1$ (equality only for $s=0$ reflecting normalization).
The following interpretation of Eq. (\ref{laplansteps}) appears mention worthy. Taking into account
the initial condition of $\Phi^{(n)}(0) =\delta_{n0}$, i.e. zero for $n \geq 1$ then it follows that
\vskip -10pt%
\begin{align}\nonumber
  \Phi^{(n)}(t) &=
 \int_0^t \left(\chi^{(n)}(\tau)-\chi^{(n+1)}(\tau)\right){\rm d}\tau \\
&= \mathcal{L}^{-1}\left\{ \frac{1}{s}\left({\tilde \chi}^{(n)}(s)-{\tilde \chi}^{(n+1)}(s)\right) \right\},  \label{cumulchin}
\end{align}
\vskip -3pt \noindent%
where $\int_0^t \chi^{(m)}(\tau){\rm d}\tau$ is the probability for
{\it at least} $m$ arrivals within $[0,t]$. It follows that
the difference in (\ref{cumulchin}) gives the probability $\Phi^{(n)}(t)$ for {\it exactly} $n$ arrivals within $[0,t]$.

It is convenient now to introduce a (dimensionless) generating function $G(t,v)$ which allows us to recover
many quantities of interest in elegant manner. We define this generating function as
\vskip - 11pt%
\begin{equation}
 \label{generatingsteps}
 G(t,v)= \sum_{n=0}^{\infty} \Phi^{(n)}(t)\, v^n ,\hspace{1cm} |v|\leq 1,
\end{equation}
\vskip -2pt \noindent %
where the value $G(t,1)=1$ reflects normalization condition (\ref{phinnormaliszation}). In the
Laplace domain we get
\vskip - 12pt%
\begin{equation}
 \label{generatingstepsLaplacetrafo}
 \begin{array}{l}
 \ds {\tilde G}(s,v) = \sum_{n=0}^{\infty} {\tilde \Phi}^{(n)}(s)v^n  =
 \frac{1-{\tilde \chi}(s)}{s}\sum_{n=0}^{\infty} ({\tilde \chi}(s))^n v^n \\  
 \ds \hspace{1.3cm} =\frac{1-{\tilde \chi}(s)}{s} \frac{1}{1-v{\tilde \chi}(s)} ,\hspace{1cm} |v|\leq 1 .
 \end{array}
\end{equation}
\vskip -3pt \noindent %
We observe that  ${\tilde G}(s,v=1)= s^{-1}$
($s\neq 0$) and thus proves $G(t,1) = \sum_{n=0}^{\infty} \Phi^{(n)}(t) = 1$, i.e.
normalization relation (\ref{phinnormaliszation}). It follows from $|{\tilde \chi}(s)| \leq {\tilde \chi}(s=0)=1$ that
the series (\ref{generatingstepsLaplacetrafo}) is converging for $|v| \leq 1$.

A quantity of great interest in the following analysis is
the expected number of arrivals ${\bar n}(t)$ (expected number of jumps) that happen in the time interval $[0,t]$. This quantity is obtained by
\vskip -11pt%
\begin{equation}
 \label{averageNumberSteps}
 \ds {\bar n}(t) = \sum_{n=0}^{\infty} n \Phi^{(n)}(t) = \frac{d}{d v} G(t,v){\Big|}_{v=1} =
 \mathcal{L}^{-1} \left\{  \frac{{\tilde \chi}(s)}{s (1-{\tilde \chi}(s))} \right\} , 
\end{equation}
\vskip -2pt \noindent %
${\bar n}(t)$ is a dimensionless function of $t$. 
In view of relation (\ref{averageNumberSteps}) the following observation appears mention worthy. 
Taking into account the identity
\vskip -10pt%
\begin{equation}
 \label{mentionobs}
 \left({\tilde {\bar n}}(s) +\frac{1}{s}\right){\tilde {\chi(s)}} = \frac{\chi(s)}{s(1-{\tilde \chi}(s))}=  {\tilde {\bar n}}(s),
\end{equation}
\vskip -1pt \noindent%
where in the time domain this relation reads
\begin{equation}
 \label{timedomainrel}
 {\bar n}(t)=\int_0^t(1+{\bar n}(\tau))\chi(t-\tau){\rm d}\tau =
 1-\Phi^{(0)}(t)+\int_0^t{\bar n}(\tau)\chi(t-\tau){\rm d}\tau,
\end{equation}
and is referred to as {\it renewal equation} \cite{Cox1967,GorenfloMainardi2013,Gorenflo2010}.
The renewal equation is equivalently obtained by
\vskip -11pt%
\begin{equation}
 \label{renewaleq}
 {\bar n}(t)= \sum_{n=0}^{\infty}(n+1)\Phi^{(n+1)}(t) =
 \sum_{n=0}^{\infty} n\Phi^{(n+1)} + \sum_{n=0}^{\infty} \Phi^{(n+1)},
\end{equation}
\vskip -3pt \noindent
where we identify $\sum_{n=0}^{\infty} n\Phi^{(n+1)} =
\int_0^t{\bar n}(\tau)\chi(t-\tau){\rm d}\tau $ and $\sum_{n=0}^{\infty} \Phi^{(n+1)} = 1-\Phi^{0}(t) = \Psi(t)$ as
the two contributions of Eq. (\ref{timedomainrel}).
\vspace*{-3pt} 
\section{Fractional Poisson distribution and process} \label{FPP}
\setcounter{section}{3} \setcounter{equation}{0} \setcounter{theorem}{0}

In this section, we recall the {\it fractional Poisson process}
introduced by Laskin who coined this name \cite{Laskin2003}. The fractional Poisson process also was derived in different manners by several authors
\cite{BeghinOrsinger2009,GorenfloMainardi2013,MainardiGorenfloScalas2004,RepinSaichev2000}.
This process is a natural generalization of
the standard Poisson process. The subsequent Section \ref{GFPP-gen} is devoted to a generalization
of Laskin's fractional Poisson process which we call the `generalized fractional Poisson process' (GFPP).

In order to capture effects of causality we employ the machinery of causal Green's functions (causal propagators)
and their relations to Laplace transforms. Let us recall some basic properties of this machinery.
We introduce the causal propagator $g_{\beta}(t)=\Theta(t)g_{\beta}(t)$ by
\vskip -10pt%
\begin{equation}
\label{FGdef}
\left(\frac{d}{dt}+\sigma\right)^{\beta} g_{\beta,\sigma}(t)= \delta(t)
,\hspace{1cm} \beta \geq 0 \hspace{1cm} \sigma > 0 .
\end{equation}
\vskip -1pt \noindent
By using $\delta(t)= (2\pi)^{-1}\int_{-\infty}^{\infty}e^{i\omega t}{\rm d}\omega$ for
the Dirac $\delta$-distribution, we obtain the Fourier transform of this propagator
as ${\bar g}_{\beta,\sigma}(\omega)=\left(\sigma + i\omega\right)^{-\beta}$. This
propagator is then defined by the Fourier integral
\vskip -10pt%
\begin{equation}
\label{FouriertrafoGF}
g_{\beta,\sigma}(t) =
\frac{1}{2\pi}\int_{-\infty}^{\infty} \frac{e^{i\omega t}}{(\sigma+i\omega)^{\beta}} {\rm d}\omega ,
\end{equation}
\vskip -3pt \noindent
where we observe that
\vskip -10pt%
\begin{equation}
\label{auxiliary}
  \frac{1}{(\sigma+i\omega)^{\beta}} =
\int_{-\infty}^{\infty}\Theta(\tau) e^{-(\sigma+i\omega)\tau}\frac{\tau^{\beta-1}}{(\beta-1)!}{\rm d}\tau.
\end{equation}
We employ throughout this paper as equivalent notations $ \zeta ! =\Gamma(\zeta+1)$.
Plugging relation (\ref{auxiliary}) into (\ref{FouriertrafoGF}) yields
\vskip -10pt%
\begin{equation}
 g_{\beta,\sigma}(t) =
 \left(\frac{d}{dt}+\sigma\right)^{-\beta}\delta(t)=  e^{-\sigma t} \Theta(t)\frac{t^{\beta-1}}{(\beta-1)!} = e^{-\sigma t}g_{\beta}(t),
\label{greensfunctionofOrderbeta}
\end{equation}
\vskip -3pt \noindent
where $\sigma >0$. With these equations we can write
\begin{align}\nonumber
 g_{\beta}(t) &= e^{\sigma t} g_{\beta,\sigma}(t) = e^{\sigma t}
 \frac{1}{2\pi}\int_{-\infty}^{\infty} \frac{e^{i\omega t}}{(\sigma+i\omega)^{\beta}} {\rm d}\omega\\
&= \Theta(t)\frac{t^{\beta-1}}{\Gamma(\beta)} =
 \mathcal{L}^{-1}\{s^{-\beta}\} =
   \frac{1}{2\pi i }\int_{\sigma-i\infty}^{\sigma +i \infty} s^{-\beta} e^{st}{\rm d}s   \label{backtrafo}
\end{align}
with $s=\sigma +i\omega$ and $\sigma=\Re(s) >0$ where the least line shows the equivalence to
Laplace inversion.

\smallskip 

We identify the causal propagator
$g_{\beta}(t)=\lim_{\sigma\rightarrow 0+} g_{\beta,\sigma}(t)= \Theta(t)\frac{t^{\beta-1}}{\Gamma(\beta)}$  with
the kernel of the Riemann-Liouville fractional integral of order $\beta$. This kernel includes the limit $\lim_{\beta\rightarrow 0+}g_{\beta}(t) =\delta(t)$.
Hence $g_{\beta}(t)$ is the causal Green's function
that solves $(\frac{d}{dt})^{\beta}g_{\beta}(t) = \delta(t)$.
Applying $(\frac{d}{dt})^{-\beta}$ on a causal function $f(t)=\Theta(t)f(t)$ hence gives
the convolution
\vskip -10pt%
\begin{align}
 \label{diffint}
\lim_{\epsilon\rightarrow 0+} &\left(\frac{d}{dt}+\epsilon\right)^{-\beta} f(t) =  \frac{1}{\Gamma(\beta)} \int_0^{t}  (t-\tau)^{\beta-1} f(\tau){\rm d}\tau
,\hspace{0.8cm} \beta \geq 0,
\end{align}
which we identify with the Riemann-Liouville fractional integral of order $\beta$
\cite{GorenfloKilbas2014,GorenfloMainardi2013,OldhamSpanier1974}. In Appendix \ref{fractionalPoissonprocess} we show
that the causal propagator $ (\frac{d}{dt})^{\beta}\delta(t)$ defines the Riemann-Liouville fractional derivative kernel.
These relations shed light on the connection of fractional operators with causal propagators.
\smallskip 

With these general remarks, let us now introduce the causal propagator (the causal waiting time PDF) that defines Laskin's {\it fractional Poisson process}
(See Appendix \ref{fractionalPoissonprocess} for some properties):
\vskip -10pt%
\begin{align}\nonumber
G_{\beta,\xi}(t) &= \sum_{n=1}^{\infty}  (-1)^{n-1} \xi^{n} g_{n\beta}(t) =
 \Theta(t)  \sum_{n=1}^{\infty}(-1)^{n-1} \xi^{n}
\frac{ t^{n\beta-1}}{(n\beta-1)!}  \\ \nonumber
&= \xi t^{\beta-1} \sum_{n=0}^{\infty}\frac{(-1)^n \xi^n t^{n\beta}}{\Gamma(\beta n+\beta)}= \xi t^{\beta-1} E_{\beta,\beta}(-\xi t^{\beta})  ,\hspace{1cm} t\in \mathbb{R}_{+} \\
&=  \frac{d}{dt}[1-E_{\beta}(-\xi t^{\beta})]
,\hspace{0.5cm}  0<\beta\leq 1  ,\hspace{0.5cm}
\label{Mittag-Leffler-waiting-time-PDF}
\end{align}
which is the Mittag-Leffler density containing the Mittag-Leffler function of order $\beta$ \cite{GorenfloKilbas2014,Gorenflo2010,Gorenflo2007,Laskin2003}
\vskip -10pt %
\begin{equation}
 \label{Mittag-L-intro}
E_{\beta}(z) = \sum_{n=0}^{\infty} \frac{z^n}{\Gamma(\beta n+1)} ,\hspace{0.8cm} \Re\{\beta\} >0 ,
\hspace{0.4cm} z \in \mathbb{C},
\end{equation}
and the generalized (two-parameter) Mittag-Leffler function 
\vskip -10pt%
\begin{equation}
 \label{Mittag-L-intro-gener}
E_{\beta,\gamma}(z) = \sum_{n=0}^{\infty} \frac{z^n}{\Gamma(\beta n+\gamma)} , \hspace{0.5cm} \Re\{\beta\} >0 ,
\hspace{0.8cm} \Re\{\gamma\} >0 , \hspace{0.4cm} z \in \mathbb{C},
\end{equation}
\vskip -3pt \noindent
where $E_{\beta,1}(z) = E_{\beta}(z)$ recovers the Mittag-Leffler function (\ref{Mittag-L-intro}) of order $\beta$.
Since the argument $-\xi t^{\beta}$ of the Mittag-Leffler density (\ref{Mittag-Leffler-waiting-time-PDF}) is dimensionless,
$\xi$ is a positive characteristic parameter of physical dimension
$[sec]^{-\beta}$.
Let us now rederive the Mittag-Leffler density in the following way establishing the connection of causal propagators with Laplace transforms
\begin{align}\nonumber
G_{\beta,\xi}(t) &= e^{\sigma t}G_{\beta,\sigma,\xi}(t)= e^{\sigma t} \sum_{n=1}^{\infty}
 \xi^{n} (-1)^{n-1} \left(\frac{d}{dt}+\sigma\right)^{-n\beta}\delta(t)\\ \nonumber
& = e^{\sigma t} \xi \left\{\xi+\left(\frac{d}{dt}+\sigma \right)^{\beta}\right\}^{-1}\delta(t) \\ \nonumber
&= \frac{e^{\sigma t}}{2\pi} \int_{-\infty}^{\infty}  {\rm d}\omega \sum_{n=1}^{\infty}
 \xi^{n} (-1)^{n-1} (\sigma+i\omega)^{-\beta n}  \\
&= \frac{e^{\sigma t}}{2\pi}
 \int_{-\infty}^{\infty} e^{i\omega t} \frac{\xi}{\xi + (\sigma +i\omega)^{\beta}} {\rm d}\omega =
 \mathcal{L}^{-1}\left\{\frac{\xi}{\xi + s^{\beta}}\right\} .
\label{rederive}
\end{align}
The convergence of the series requires $\sigma =\Re\{s\} > \xi^{1/\beta}$, where
$s=\sigma+i\omega$ can be identified with the Laplace variable.
We observe then
\vskip -10pt%
\begin{equation}
 \label{additionalGener}
G_{\beta,\sigma,\xi}(t)= e^ {-\sigma t}G_{\beta,\xi}(t) =\sum_{n=0}^{\infty}(-1)^{n-1}\xi^n g_{n\beta,\sigma}(t),
\end{equation}
\vskip -2pt \noindent
where the causal propagators $G_{\beta,\sigma,\xi}(t)$, and $G_{\beta,\xi}(t)$ fulfill
\begin{align}
\nonumber
 \left\{\xi+\left(\frac{d}{dt}+\sigma \right)^{\beta}\right\} G_{\beta,\sigma,\xi}(t) &= \xi \delta(t),\\
\left\{\xi+\left(\frac{d}{dt}\right)^{\beta}\right\} G_{\beta,\xi}(t) &= \xi \delta(t).\label{solvesequation}
\end{align}
One can then introduce the fractional Poisson process as the renewal process with
the Mittag-Leffler
jump density (\ref{Mittag-Leffler-waiting-time-PDF}) which reduces for
$\beta=1$ to the exponential density $\chi_1(t)=\xi e^{-\xi t} $ of the standard Poisson process.

The survival probability in the fractional Poisson process
is of the form of a Mittag-Leffler function
\vskip -10pt %
\begin{align}
\nonumber
&\Phi^{(0)}_{\beta}(t)= 1-\Psi_{\beta}(t) = \mathcal{L}^{-1}\left\{\frac{s^{\beta-1}}{(\xi+s^{\beta})} \right\}\\
&= \int_t^{\infty} G_{\beta,\xi}(\tau){\rm d}\tau =E_{\beta}(-\xi t^{\beta})
= \sum_{n=0}^{\infty} \frac{(-\xi t^{\beta})^n}{\Gamma(n\beta+1)}
,\hspace{0.4cm} t\geq 0, 
\label{survival}
\end{align}
\vskip -2pt \noindent
where for $\beta=1$ the standard Poisson process with survival probability $E_1(-\xi t) = e^{-\xi t}$ is recovered.
The {\it fractional Poisson distribution} is defined by
the probabilities $\Phi_{\beta}^{(n)}(t)$ for $n$ arrivals within $[0,t]$
with jump density (\ref{Mittag-Leffler-waiting-time-PDF}) and is
obtained as \cite{Laskin2003} (See Appendix \ref{fractionalPoissonprocess} for a derivation):
\begin{align}
 \nonumber
& \Phi^{(n)}_{\beta}(t) = \frac{(\xi t^{\beta})^n}{n!}\frac{{\rm d}^n}{{\rm d}\tau^n}E_{\beta}(\tau){\Big|}_{\tau=-\xi t^{\beta}} ,\hspace{0.8cm} t \in \mathbb{R}_{+} \\
& =  \frac{(\xi t^{\beta})^n}{n!} \sum_{m=0}^{\infty}\frac{(m+n)!}{m!}\frac{(-\xi t^{\beta})^m}{\Gamma(\beta(m+n)+1)}, \qquad n=0,1,2,\ldots,  \label{fractionalpoission-distribution}
\end{align}
\vskip -1pt \noindent
where $0<\beta \leq 1$.
For $\beta=1$ the fractional Poisson distribution recovers
the standard Poisson distribution $\Phi^{(n)}_{1}(t)=\frac{(\xi t)^n}{n!}e^{-\xi t}$.
%
\section{Generalization of the fractional Poisson process} \label{GFPP-gen}

\setcounter{section}{4} \setcounter{equation}{0} \setcounter{theorem}{0} 
\subsection{Jump density and survival probability distribution} 
\label{generalized}
Here we survey the `{\it Generalized Fractional Poisson Process}' ({\it GFPP}) first introduced by Cahoy and Polito \cite{CahoyPolito2013}. The GFPP is a generalization of the Laskin fractional Poisson process.
The jump density (waiting time PDF) of the GFPP has the Laplace transform \cite{CahoyPolito2013,MichelitschRiascosPhysA2020}
\begin{equation}
 \label{jump-gen-fractional-laplacetr}
 {\tilde \chi}_{\beta,\alpha}(s) = \frac{\xi^{\alpha}}{(s^{\beta}+\xi)^{\alpha}} ,\hspace{0.5cm}
 0<\beta\leq 1 ,\hspace{0.5cm} \alpha >0 , \hspace{1cm} \xi >0.
\end{equation}
The characteristic dimensional constant $\xi$ in
(\ref{jump-gen-fractional-laplacetr}) has physical dimension $[sec]^{-\beta}$.
Per construction
${\tilde \chi}_{\beta,\alpha}(s){\big|}_{s=0}= \int_0^{\infty} \chi_{\beta,\alpha}(t){\rm d}t=1$
reflects normalization of the jump density.
The GFPP recovers for $\alpha=1$ with $0<\beta\leq 1$ the fractional Poisson process, for $\beta=1$ with $\alpha>0$
an Erlang type process where we allow any $\alpha >0$, i.e.
also non-integer $\alpha$ \cite{GorenfloMainardi2013},
and for $\alpha=1$, $\beta=1$ the standard Poisson process.
For simplicity we call subsequently the process with $\beta=1$, $\alpha >0$ (for any $\alpha \in \mathbb{R}_{+}$)
 `Erlang process'.

\begin{figure}
\begin{center}
\includegraphics[width=1.0\textwidth]{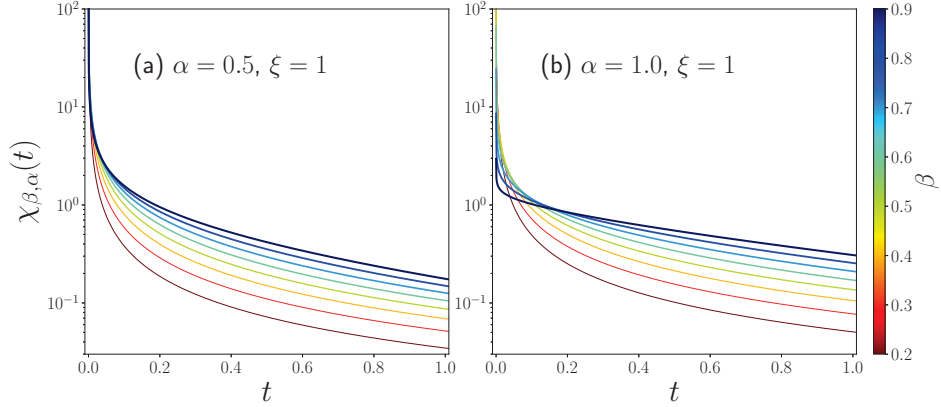}
\end{center}
\caption{\label{Figure1} 
Prabhakar waiting time PDF $\chi_{\beta,\alpha}(t)$ of Eq. (\ref{Laplainv}) for (a) $\alpha=0.5$ and
(b) $\alpha=1.0$ for different values  $0< \beta \leq 1$.
The case  $\alpha=1$  of Eq. (\ref{Laplainv})
recovers the Mittag-Leffler density of the Laskin fractional Poisson process
(Eq. (\ref{Mittag-Leffler-waiting-time-PDF})) for different values  $0< \beta \leq 1$. The results were obtained numerically using $\xi=1$. }
\end{figure}

\begin{figure}
\begin{center}
\includegraphics[width=1.0\textwidth]{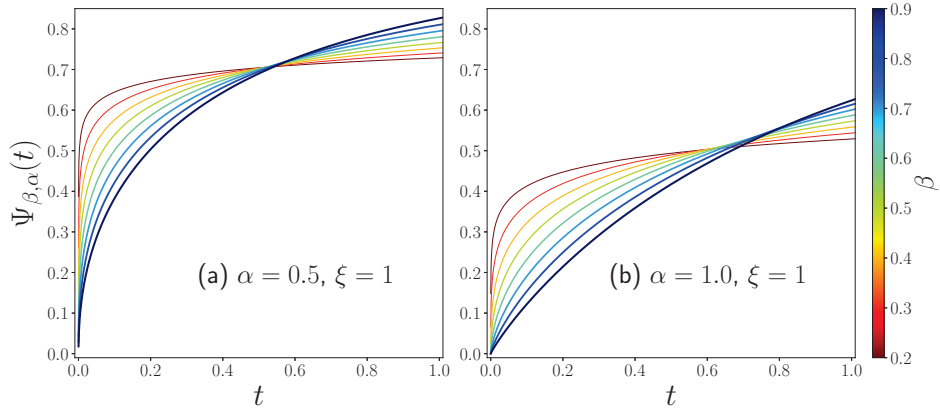}
\end{center}
\caption{\label{Figure2} 
Cumulative probability $\Psi_{\beta,\alpha}(t)$ of Eq. (\ref{cumulgenfract})
for (a) $\alpha=0.5$ and (b) $\alpha=1.0$ for different values of $\beta$ in the interval
$0< \beta \leq 1$. The case with $\alpha=1$ reduces to the failure probability $\Psi_{\beta,\alpha=1}(t) = 1-E_{\beta}(-\xi t^{\beta})$ (recovered by Eq. (\ref{cumulgenfract}) for $\alpha=1$)
of the Laskin fractional Poisson process for different values of $0< \beta \leq 1$.
The results were obtained numerically with $\xi=1$.}
\end{figure}

We observe in (\ref{jump-gen-fractional-laplacetr}) that the jump density
in the fractional range $0<\beta<1$ and for all
$\alpha>0$ has
diverging mean (diverging expected time of first arrival), namely
$-\frac{d}{ds}{\tilde \chi}_{\beta,\alpha}(s){\big|}_{s=0}=\int_0^{\infty} t
\chi_{\beta,\alpha}(t) \rightarrow +\infty$ reflecting occurrence of very long waiting times.
In order to determine the time domain representation of (\ref{jump-gen-fractional-laplacetr}) it is convenient
to introduce the Pochhammer symbol which is defined as \cite{Mathai2010}
\begin{equation}
\label{Pochhammer}
(c)_m  =\frac{\Gamma(c+m)}{\Gamma(c)} = \left\{\begin{array}{l} 1 ,\hspace{5mm} m=0 \\ \\

     c (c+1)\ldots (c+m-1) ,\hspace{5mm} m=1,2,\ldots. \end{array}\right.
\end{equation}
For $c \neq 0$ (\ref{Pochhammer})
can be represented as $(c)_m = \frac{\Gamma(c+m)}{\Gamma(c)}$.
For $c=0$ this expression has be regularized to define its value at $c=0$ with
$(0)_m=\delta_{m0} =\lim_{c\rightarrow 0+}\frac{\Gamma(c+m)}{\Gamma(c)}$
which occurs as an important limiting case
for $\lim_{\alpha\rightarrow 0+}\chi_{\beta,\alpha}(t) =\mathcal{L}^{-1} (1)=\delta(t)$.
With this regularization definition (\ref{Pochhammer}) can be extended to all $c\in \mathbb{C}$.
We can then evaluate the jump density defined by (\ref{jump-gen-fractional-laplacetr}) by taking into account
($0<\beta\leq 1$, $\alpha >0$)
\begin{equation}
\label{simaliarlyasmittag-leffler}
\begin{array}{l}
\ds {\tilde \chi}_{\beta,\alpha}(s) =
\frac{\xi^{\alpha} s^{-\beta\alpha }}{(1+\xi s^{-\beta})^{\alpha}}
= (\xi s^{-\beta})^{\alpha} \sum_{m=0}^{\infty}
(-1)^m \frac{\Gamma(\alpha+m)}{m!\Gamma(\alpha)} \xi^m s^{-\beta m}   \\  
\ds \hspace{0.5cm} = \xi^{\alpha}
\sum_{m=0}^{\infty} (-1)^m \frac{(\alpha)_m}{m!} \xi^m s^{-\beta (m+\alpha)}
,\hspace{1cm}  \sigma=\Re(s) >\xi^{\frac{1}{\beta}}   .
\end{array}
\end{equation}
In the same way as above (See Eq. (\ref{rederive})) we get the causal time domain representation of the jump density of the GFPP in terms of causal propagators in form \cite{CahoyPolito2013,MichelitschRiascosPhysA2020}
\begin{align}
\nonumber
\chi_{\beta,\alpha}(t) &= \mathcal{L}^{-1}\left\{  \frac{\xi^{\alpha}}{(s^{\beta}+\xi)^{\alpha}} \right\}= e^{\sigma t}\xi^{\alpha}\left(\xi+(\sigma+\frac{d}{dt})^{\beta}\right)^{-\alpha}\delta(t) \\ \nonumber
&= \frac{e^{\sigma t}}{2\pi}
 \int_{-\infty}^{\infty} e^{i\omega t} \frac{\xi^{\alpha}}{(\xi + (\sigma +i\omega)^{\beta})^{\alpha}} {\rm d}\omega
 \\ \nonumber
&= \xi^{\alpha} \sum_{m=0}^{\infty} (-1)^m \frac{(\alpha)_m}{m!}
 \xi^m \mathcal{L}^{-1}\{s^{-\beta (m+\alpha)}\} ,\hspace{0.5cm} \sigma =\Re\{s\} >\xi^{\frac{1}{\beta}}
 \\ \nonumber
&= \xi^{\alpha} \frac{t^{\beta\alpha-1}}{\Gamma(\alpha\beta)}+ \xi^{\alpha} t^{\beta\alpha-1}
 \sum_{m=1}^{\infty}\frac{(\alpha)_m}{m!}\frac{(-\xi t^{\beta})^m}{\Gamma(\beta m+ \alpha\beta)}
 \\
& =  \xi^{\alpha} t^{\beta\alpha-1} E_{\beta,(\alpha\beta)}^{\alpha}(-\xi t^{\beta}) ,\hspace{1cm} 0<\beta\leq 1, \hspace{0.5cm} \alpha >0.
 \label{Laplainv}
 \end{align}
For $\alpha=1$ (\ref{Laplainv}) reduces to the Mittag-Leffler density (\ref{Mittag-Leffler-waiting-time-PDF}), then for $\beta=1$ with $\alpha>0$ to the so called Erlang-density
$\chi_{1,\alpha}(t)= \frac{\xi^{\alpha} t^{\alpha-1}}{\Gamma(\alpha)}e^{-\xi t}$ and
for $\alpha=1$,
$\beta=1$ to the density $\chi_{1,1}(t)=\xi e^{-\xi t}$ of the standard Poisson process.
In expression (\ref{Laplainv}) appears a generalization of the Mittag-Leffler
function, the so called Prabhakar function which was introduced by Prabhakar \cite{Prabhakar1971}
and has the representation
\begin{equation}
 \label{genmittag-Leff}
 E_{a,b}^c(z) = \sum_{m=0}^{\infty}
 \frac{(c)_m}{m!}\frac{z^m}{\Gamma(am + b)} ,\hspace{0.25cm} \Re(a) >0, \hspace{0.25cm}
 a, b, c, z \in \mathbb{C} .
\end{equation}
The Prabhakar function converges for all $z\in {\mathbb{C}}$.
The Prabhakar function is analyzed by several authors \cite{GarraGarrappa2018,Mathai2010,ShulkaPrajabati2007}
and has recently attracted much interest. We call (\ref{Laplainv}) `Prabhakar density' or
also `GFPP (jump-) density'.
The GFPP (Prabhakar generalization of fractional Poisson counting process) and the related Prabhakar generalization of fractional calculus is a newly emerging rapidly developing field.
For a comprehensive
review of properties and applications we refer to the recent review article by Giusti et al.
\cite{Giusti-et-al2020} and for an outline of Prabhakar theory and related generalized fractional calculus consult
\cite{GarraGorenfloPolito2014,Giusti2020,PolitoTomovski2016}.
For an outline of Mittag-Leffler functions of several types and related topics we refer
to the references
\cite{HauboldMathaiSaxena2011,KilbasSrivastavaTruillo2006,MillerRoss1993,Podlubny1999,SamkoKilbasMarichev1993}.
Physical applications of the Prabhakar theory can be found in the references
\cite{dosSantos2019,GiustiColombaro2018,MichelitschRiascos-etal2020,Sandev2017} and an analysis of complete monotonicity of the Prabhakar function
with a non-Debye relaxation model is developed in \cite{MainardiGarrappa2015}. Discrete-time variants of Prabhakar renewal processes are developed in a recent paper \cite{MichelitschPolitoRiascos2020}.
We observe that for $t$ small the Prabhakar density (\ref{Laplainv}) behaves as
\begin{equation}
 \label{lowest-order-in-t}
 \lim_{t \rightarrow 0} \chi_{\beta,\alpha}(t)= \lim_{t \rightarrow 0} \frac{\xi^{\alpha} t^{\alpha\beta-1}  }{\Gamma(\alpha\beta)}
\rightarrow  \left\{\begin{array}{l} 0    ,\hspace{1cm}    \alpha\beta>1, \\ \\
 \frac{\xi^{\alpha}}{\Gamma(\alpha\beta)} , \hspace{1cm} \alpha\beta=1, \\ \\
         \infty \hspace{1cm}   \alpha\beta <1.
        \end{array} \right.
\end{equation}
The power-law asymptotic behavior $\sim t^{\alpha\beta-1}$ occurring for $t$ small
is different from that of the fractional Poisson process
which is weakly singular $\sim t^{\beta-1}$ and is reproduced for $\alpha=1$.
The Prabhakar density is weakly singular only for $\alpha\beta < 1$ and non-singular else.
The Prabhakar density (\ref{Laplainv}) is plotted in Fig. \ref{Figure1} for different values
$0<\beta<1$ and $\alpha=0.5$, $\alpha=1$, respectively. $\alpha=1$ in Fig. \ref{Figure1}(b) recovers
the Mittag-Leffler density (\ref{Mittag-Leffler-waiting-time-PDF}). We see in this plot the behavior
$\lim_{t\rightarrow 0}\chi_{\beta,\alpha}(t) \sim t^{\alpha\beta-1} \rightarrow +\infty$  occurring for $\alpha\beta <1$. The jump frequency in this regime is extremely
high for short observation times where the jump density takes $\delta(t)$-shape in the limit
$\alpha\beta\rightarrow 0+$.
We can directly integrate (\ref{Laplainv}) to obtain the cumulative distribution
defined in Eq.
(\ref{waitintimedis}), namely
\vskip -12pt %
\begin{align}
\nonumber
&\Psi_{\beta,\alpha}(t)= \int_0^{t}  \chi_{\beta,\alpha}(\tau){\rm d}\tau = \xi^{\alpha} \sum_{n=0}^{\infty}
(-1)^n \frac{(\alpha)_n}{n!}
 \frac{\xi^n t^{\beta (n+\alpha)}}{(\beta(n+\alpha))!}\\
 &=
 \xi^{\alpha} t^{\alpha\beta} \sum_{n=0}^{\infty}\frac{(\alpha)_n}{n!}
 \frac{(-\xi t^{\beta})^n}{\Gamma(\beta n + \alpha\beta+1)} =
 \xi^{\alpha} t^{\alpha\beta} E_{\beta,(\alpha\beta+1)}^{\alpha}(-\xi t^{\beta}) .
\label{cumulgenfract}
\end{align}
\vskip -2pt \noindent
The probability distribution (\ref{cumulgenfract}) (failure probability) of at least one arrival within $[0,t]$ in a GFPP
is plotted for different values of
$0<\beta<1$ in Fig. \ref{Figure2} for $\alpha=0.5$ and $\alpha=1$ (Mittag-Leffler case),
respectively. For small $\xi^{\frac{1}{\beta}}t$ this distribution is dominated by the order
$n=0$, namely
\vskip - 10pt%
\begin{equation}
 \label{smalldmt}
 \Psi_{\beta,\alpha}(t)
 \approx \frac{\xi^{\alpha}}{\Gamma(\alpha\beta+1)} t^{\alpha\beta} ,\hspace{1cm}
 \xi^{\frac{1}{\beta}}t \rightarrow 0 ,
\end{equation}
\vskip -2pt \noindent
where we also see in this relation that initial
condition $\Psi_{\beta,\alpha}(0)=0$ is always fulfilled.
The $t^{\alpha\beta}$ power-law behavior can be observed in the Fig. \ref{Figure2}.
The smaller $\alpha\beta \rightarrow 0$ the more $\Psi_{\beta,\alpha}(t)$ approaches Heaviside step
function shape.
The GFPP survival probability yields
\begin{equation}
 \label{survivalgenfractpro}
 \Phi_{\beta,\alpha}^{(0)}(t)= \int_t^{\infty}\chi_{\beta,\alpha}(\tau){\rm d}\tau  =1-\Psi_{\beta,\alpha}(t)=
 1- \xi^{\alpha} t^{\alpha\beta} E_{\beta,(\alpha\beta+1)}^{\alpha}(-\xi t^{\beta})
\end{equation}
with initial condition $\Phi_{\beta,\alpha}^{(0)}(0)=1$ and
$\lim_{t\rightarrow\infty} \Phi_{\beta,\alpha}^{(0)}(t)=0$. For $\alpha=1$ (\ref{survivalgenfractpro}) becomes
\vskip - 12pt%
\begin{align}
\nonumber
\Phi_{\beta,1}^{(0)}(t) &= 1- \xi t^{\beta} E_{\beta,(\beta+1)}^{1}(-\xi t^{\beta}) \\ \nonumber
&= 1+\sum_{n=0}^{\infty}\frac{(-\xi t^{\beta})^{n+1}}{\Gamma(\beta n+\beta+1)}\\
& = \sum_{n=0}^{\infty} \frac{(-\xi t^{\beta})^n}{\Gamma(\beta n+1)} = E_{\beta}(-\xi t^{\beta}),
 \label{fractPoisson}
\end{align}
\vskip -3pt \noindent
reproducing the Mittag-Leffler survival probability (\ref{survival}) in the fractional Poisson process.
On the other hand let us consider $\beta=1$ and $\alpha >0$ which gives
\vskip -13pt 
\begin{equation}
 \label{beta1alphachi}
 \chi_{1,\alpha}(t) = \xi^{\alpha} t^{\alpha-1}E^{\alpha}_{1,\alpha}(-\xi t) = \xi^{\alpha}\frac{t^{\alpha-1}}{\Gamma(\alpha)} \sum_{n=0}^{\infty}\frac{(-\xi t)^n}{n!} =
 \xi^{\alpha}\frac{t^{\alpha-1}}{\Gamma(\alpha)} e^{-\xi t},
\end{equation}
\vskip -2pt \noindent
i.e. for $\beta=1$ the jump density is light-tailed with exponentially evanescent behavior for $t\rightarrow \infty$, but for $t\rightarrow 0$
we get power law behavior $\chi_{1,\alpha}(t) \sim t^{\alpha-1}$.
The density (\ref{beta1alphachi}) is an Erlang type density
(extending the traditional Erlang density to any $\alpha \in \mathbb{R}_{+}$).
For $\alpha=1$, $\beta=1$ (\ref{beta1alphachi}) recovers
the density $\chi_{1,1}(t)= \xi e^{-\xi t}$ of the standard Poisson process.
Let us now consider the behavior for large times $t\rightarrow \infty$.
Expanding (\ref{jump-gen-fractional-laplacetr}) for $|s|$ small yields
\vskip -11pt%
\begin{equation}
 \label{jump-gen-fractional}
 {\tilde \chi}_{\beta,\alpha}(s) = \left(1+\frac{s^{\beta}}{\xi}\right)^{-\alpha} = \sum_{n=0}^{\infty}
 \frac{(\alpha)_n}{n!}(-1)^n \xi^{-n}s^{n\beta} = 1-\frac{\alpha}{\xi} s^{\beta}+\ldots .
 \end{equation}
\vskip -3pt \noindent
Thus we get fat-tailed behavior, namely
\vskip -10pt%
 \begin{equation}
  \label{tlarge-fat-tailed}
  \chi_{\beta,\alpha}(t) \approx -\frac{\alpha}{\xi \Gamma(-\beta)}t^{-\beta-1} ,
  \hspace{0.5cm} 0<\beta< 1 ,\hspace{0.5cm} \alpha>0 ,\hspace{0.5cm}
  t\rightarrow\infty ,
 \end{equation}
where the exponent is independent of $\alpha$ and with
$-\Gamma(-\beta)=\frac{\Gamma(1-\beta)}{\beta} >0$ for $0<\beta<1$.

\subsection{Generalized fractional Poisson distribution} 
\label{GeneralizedFractPoissonDistri}

In this section we deduce the generalized counterpart $\Phi^{(n)}_{\beta,\alpha}(t)$ to the fractional Poisson distribution (\ref{fractionalpoission-distribution}).
The probability distribution of $n$ arrivals within $[0,t]$ is with  (\ref{laplansteps}) and
(\ref{jump-gen-fractional-laplacetr}) defined by
\begin{align}
\nonumber
\Phi^{(n)}_{\beta,\alpha}(t) &=\mathcal{L}^{-1}\left\{
\frac{1}{s}\left({\tilde \chi}^n_{\beta,\alpha}(s)-{\tilde \chi}^{n+1}_{\beta,\alpha}(s)\right)\right\}
\\ \nonumber
&=\mathcal{L}^{-1}\left\{
\frac{1}{s}\left({\tilde \chi}_{\beta,n\alpha}(s)-{\tilde \chi}_{\beta,(n+1)\alpha}(s)\right) \right\} \\
&=\Psi_{\beta,n\alpha}(t)-\Psi_{\beta,(n+1)\alpha}(t)
,\hspace{1cm} n=0,1,2,\ldots,
\label{generalizeedFractPoissonDis}
\end{align}
where ${\tilde \chi}_{\beta,n\alpha}(s) =\frac{\xi^{n\alpha}}{(\xi+s^{\beta})^{n\alpha}}$ and with the
cumulative probabilities $\Psi_{\beta,n\alpha}(t)$ (Eq. (\ref{cumulgenfract})
with $\alpha\rightarrow n\alpha$). We obtain
\vskip -10pt %
\begin{align}
\nonumber
 \Psi_{\beta,n\alpha}(t) = \mathcal{L}^{-1}&\left\{\frac{1}{s}{\tilde \chi}_{\beta,n\alpha}(s)\right\} =
 \xi^{n\alpha} \sum_{m=0}^{\infty} (-1)^m \frac{(n\alpha)_m}{m!}
 \xi^m \mathcal{L}^{-1}\{s^{-\beta m- n\alpha\beta -1)}\}  \\ \nonumber
& =  \xi^{n\alpha} \sum_{m=0}^{\infty}  \frac{(n\alpha)_m}{m!}
 \frac{(-1)^m \xi^m t^{m\beta+n\alpha\beta}}{\Gamma(m\beta+n\alpha\beta+1)} \\
&=  \xi^{n\alpha} t^{n\alpha\beta} E^{n\alpha}_{\beta,(n\alpha\beta+1)}(-\xi t^{\beta}) ,  \hspace{10mm} n=0,1,2,\ldots\, .
 \label{chnprim}
\end{align}
%
%
%
With this result we obtain from (\ref{generalizeedFractPoissonDis}) the probability of $n$ arrivals within $[0,t]$ (GFPP state probabilities)
\vskip - 12pt %
\begin{multline}
 \label{Generalized-Fractional-Poisson-Distribution}
   \Phi^{(n)}_{\beta,\alpha}(t)  =
   (\xi t^{\beta})^{n\alpha} \left( E^{n\alpha}_{\beta,(n\alpha\beta+1)}(-\xi t^{\beta}) \right. \\
\left. - (\xi t^{\beta})^{\alpha} E^{(n+1)\alpha}_{\beta,((n+1)\alpha\beta+1)}(-\xi t^{\beta}) \right)
\end{multline}
($n=0,1,2,\ldots$)
which is a function of the `dimensionless time' $\xi^{\frac{1}{\beta}}t$ and contains the two
index parameters $0<\beta\leq 1$ and $\alpha>0$. Expression (\ref{Generalized-Fractional-Poisson-Distribution}) for the GFPP state probabilities also was obtained in \cite{CahoyPolito2013}.
We call the distribution $\Phi^{(n)}_{\beta,\alpha}(t)$ (\ref{Generalized-Fractional-Poisson-Distribution}) {\it generalized fractional Poisson distribution (GFPD)}.
For $n=0$ (\ref{Generalized-Fractional-Poisson-Distribution}) yields
$\Phi^{(0)}_{\beta,\alpha}(t)=1-\Psi_{\beta,\alpha}(t)$ of Eq. (\ref{survivalgenfractpro}).
We can write the GFPD compactly as
\vskip - 10pt%
\begin{equation}
 \label{Gen-Frac-Poisson}
 \Phi^{(n)}_{\beta,\alpha}(t) =  (\xi t^{\beta})^{n\alpha}\left\{\frac{1}{\Gamma(n\alpha\beta +1)}+
 \sum_{m=1}^{\infty} A_{\alpha,\beta}^{m,n}(t^{\beta}\xi) (-\xi t^{\beta})^m \right\}
\end{equation}
\vskip -4pt \noindent
with
\vskip - 13pt %
\begin{multline}
A_{\alpha,\beta}^{m,n}(t^{\beta}\xi) =  \frac{(n\alpha)_m}{m!\Gamma(n\alpha\beta+\beta m+1)}\\
+ \frac{(\xi t^{\beta})^{\alpha-1}\,((n+1)\alpha)_{m-1}}{(m-1)!\Gamma(n\alpha\beta+\beta m +(\alpha-1)\beta +1)},
\label{coefficient-time-dep}
\end{multline}
\vskip -3pt \noindent
for $ m=1,2,\ldots$, $n=0,1,2,\ldots$. The GFPD is a dimensionless probability distribution
where $\xi^{-\frac{1}{\beta}}$ having physical dimension $[sec]$ defines a characteristic time scale in
the GFPP.
It follows that for small dimensionless times the GFPD behaves as ($n=0,1,2,\ldots$)
\vskip -12pt %
\begin{equation}
 \label{limitsmalltilmes}
 \Phi^{(n)}_{\beta,\alpha}(t) \approx \mathcal{L}^{-1}\left(\xi^{n\alpha}s^{-n\alpha\beta -1}\right) = \frac{\xi^{n\alpha} t^{n\alpha\beta}} {\Gamma(n\alpha\beta +1)} ,\hspace{1cm}
 t\xi^{\frac{1}{\beta}} \rightarrow 0,
\end{equation}
\vskip -2pt \noindent
covering order $m=0$ of representations (\ref{Generalized-Fractional-Poisson-Distribution}) and (\ref{Gen-Frac-Poisson}).
It follows that the GFPP state probabilities fulfill the initial conditions
\vskip -10pt%
\begin{equation}
 \label{initialcon}
 \Phi^{(n)}_{\beta,\alpha}(t){\Big|}_{t=0} = \delta_{n0} ,
\end{equation}
\vskip -1pt \noindent
i.e. at $t=0$ per construction of the CTRW the walker is on his departure site.
Further of interest is the asymptotic behavior for large $t\xi^{\frac{1}{\beta}}$.
To see this behavior we expand the Laplace transform (\ref{generalizeedFractPoissonDis}) up to the
lowest non-vanishing order in $\frac{s^{\beta}}{\xi}$ to arrive at
\vskip -12pt 
\begin{equation}
\label{largetimes}
\Phi^{(n)}_{\beta,\alpha}(t) \approx  \frac{\alpha}{\xi} \mathcal{L}^{-1}\{s^{\beta-1}\}= \frac{\alpha}{\xi}
\frac{t^{-\beta}}{\Gamma(1-\beta)} ,\hspace{0.8cm} t\xi^{\frac{1}{\beta}} \rightarrow \infty,\,
\end{equation}
\vskip -2pt \noindent
where $0<\beta<1$, $\alpha >0$, $\forall n=0,1,\ldots ,\infty$. This power law holds universally for all $\alpha>0$ with $0< \beta < 1$
and contains the Laskin case $\alpha=1$ and is also independent of $n$. The universal power law scaling in
(\ref{largetimes}) indeed can be attributed to non-Markovianity and long-time memory \cite{MichelitschPolitoRiascos2020}.
We will come back to this important issue in Sect. \ref{GenFractionalPoissonProcess}.

Now let us consider the important limit $\alpha=1$ for the GFPD in more details.
Then the functions (\ref{coefficient-time-dep}) become time independent coefficients, namely
\vskip - 13pt%
\begin{align}
\nonumber
  A_{1,\beta}^{m,n} &= \frac{(n+m-1)!}{\Gamma(\beta(m+n)+1)}\left\{\frac{1}{m!(n-1)!}+\frac{1}{(m-1)!n!}\right\} \\
  &= \frac{(n+m)!}{n!m!}
  \frac{1}{\Gamma(\beta(m+n)+1)},
   \label{coefficientsalpsha1}
\end{align}
\vskip -2pt \noindent
and for the oder $m=0$ we have $\frac{1}{\Gamma(n\beta+1)}= \left(\frac{(n+m)!}{n!m!}
  \frac{1}{\Gamma(\beta(m+n)+1)}\right){\big|}_{m=0} $ thus we get for the GFPD (\ref{Gen-Frac-Poisson}) for $\alpha=1$
  the expression
\vskip -10pt%
\begin{equation}
 \label{Laskin-case}
 \Phi^{(n)}_{\beta,1}(t) = \frac{\xi^nt^{n\beta}}{n!}\sum_{m=0}^{\infty}\frac{(n+m)!}{m!}
  \frac{(-\xi)^m t^{m\beta}}{\Gamma(\beta(m+n)+1)},
\end{equation}
\vskip -2pt \noindent
which we identify with Laskin's fractional Poisson distribution (\ref{fractionalpoission-distribution}) \cite{Laskin2003}.
%
%
%
%
%
%

\subsection{Expected number of arrivals in a GFPP} 
\label{GenFractionalPoissonProcess}

Here we analyze the asymptotic behavior of the average number of arrivals ${\bar n}(t)$ within the
interval of observation $[0,t]$.
To this end we consider the generating function (\ref{generatingstepsLaplacetrafo})
\vskip - 13pt %
\begin{equation}
 \label{generatinggenfractalfunction}
 \mathcal{G}_{\beta,\alpha}(v,s) = \sum_{n=0}^{\infty} v^n {\tilde \Phi}^{(n)}_{\beta,\alpha}(s) =
 \frac{1-{\tilde \chi}_{\beta,(\alpha)}(s)}{s(1-v {\tilde \chi}_{\beta,(\alpha)}(s)} =
 \frac{1}{s} \frac{(1+\frac{s^{\beta}}{\xi})^{\alpha}-1}{(1+\frac{s^{\beta}}{\xi})^{\alpha}-v},
\end{equation}
\vskip -3pt \noindent
where $|v|\leq 1$.
The expected number ${\bar n}_{\beta,\alpha}(t)$ of steps the walker makes in the interval
$[0,t]$ is ${\bar n}_{\beta,\alpha}(t)= \frac{d}{dv}\mathcal{G}_{\beta,\alpha}(v,t){\big|}_{v=1}= \sum_{n=0}^{\infty}n\Phi^{(n)}_{\beta,\alpha}(t)$ and
has the Laplace transform
\vskip - 12pt %
\begin{equation}
 \label{Laplacetranfostepsaverge}
 {\tilde {\bar n}}_{\beta,\alpha}(s) =
 \frac{d}{dv} \mathcal{G}_{\beta,\alpha}(v,s){\Big|}_{v=1}=
 \frac{{\tilde \chi}_{\beta,\alpha}(s)}{s (1-{\tilde \chi}_{\beta,\alpha}(s))} =
 \frac{1}{s[(\frac{s^{\beta}}{\xi}+1)^{\alpha}-1]}
\end{equation}
\vskip -3pt \noindent
with ${\tilde {\bar n}}_{\beta,\alpha}(s) \approx \frac{\xi}{\alpha}s^{-\beta-1} $ for $|s|\rightarrow 0 $.
We obtain hence the asymptotic behavior for
$t\xi^{\frac{1}{\beta}} \rightarrow \infty$ large as
\vskip - 11pt%
\begin{equation}
 \label{largetimesteps}
 {\bar n}_{\beta,\alpha}(t)
\approx \frac{\xi t^{\beta} }{\alpha \Gamma(\beta+1)}
,\hspace{0.8cm} 0<\beta \leq 1 ,\hspace{3mm} \alpha >0, \hspace{3mm} t\xi^{\frac{1}{\beta}}\rightarrow\infty.
\end{equation}
\vskip -3pt \noindent
This result includes the Erlang case $\beta=1$ where for all $\alpha >0$ the average number
of arrivals ${\bar n}_{1,\alpha}(t) \approx \frac{\xi}{\alpha} t $ increases linearly for $t\xi$ large.
Now the following observation appears mention worthy. In view of the asymptotic power law behavior of the GFPD (\ref{largetimes}) which is {\it independent of $n$} and the power law (\ref{largetimesteps})
we can define
\vskip -12pt%
\begin{align}
 \nonumber
 \mathcal{C}_{\beta} &= \sum_{n=0}^{{\bar n}_{\beta,\alpha}(t)} \Phi^{(n)}_{\beta,\alpha}(t) \approx {\bar n}_{\beta,\alpha}(t) \Phi^{(0)}_{\beta,\alpha}(t)  < \sum_{n=0}^{\infty} \Phi^{(n)}_{\beta,\alpha}(t) = 1 \\ \nonumber
&= \frac{\alpha}{\xi}
\frac{t^{-\beta}}{\Gamma(1-\beta)} \frac{\xi t^{\beta}}{\alpha \Gamma(\beta+1)}   ,\hspace{0.5cm} \xi^{\frac{1}{\beta}}t \rightarrow \infty \\ \nonumber
&= \frac{1}{\Gamma(\beta+1)\Gamma(1-\beta)}\\
&= \frac{\sin{(\pi \beta)}}{\pi \beta} < 1 ,\hspace{0.5cm} 0<\beta<1 ,\hspace{3mm} \alpha >0 ,
 \label{consider}
\end{align}
\vskip -3pt \noindent
which gives for $\xi^{\frac{1}{\beta}}t \rightarrow \infty$ a universal dimensionless positive constant
$\mathcal{C}_{\beta} < 1$ which is independent of $\alpha$.
In this relation we utilized the Euler reflection formula $\Gamma(1-\beta)= \frac{\pi}{\Gamma(\beta)\sin(\pi\beta)}$.
On the other hand we obtain the asymptotic behavior for $t\xi^{\frac{1}{\beta}} \rightarrow 0 $
when we expand ${\tilde {\bar n}}_{\beta,\alpha}(s)$ for $|s|\rightarrow \infty$
to arrive at
\vskip -10pt %
\begin{align}
\nonumber
 {\bar n}_{\beta,\alpha}(t) &= \mathcal{L}^{-1}\left\{\frac{1}{s[(\frac{s^{\beta}}{\xi}+1)^{\alpha}-1]}\right\} \\
 &\approx \mathcal{L}^{-1}\left\{\xi^{\alpha}s^{-\alpha\beta-1}\right\} =
 \frac{\xi^{\alpha} t^{\alpha\beta}}{\Gamma(\alpha\beta+1)} ,\hspace{0.8cm} t\xi^{\frac{1}{\beta}}\rightarrow 0.
  \label{deductionoftsmall}
\end{align}
\vskip -3pt \noindent
Both asymptotic expressions
(\ref{largetimesteps}) and (\ref{deductionoftsmall}) coincide for $\alpha=1$
recovering the (exact) expression (\ref{averagefractionalsteps}) for the fractional Poisson process.
%
%

\section{Montroll-Weiss CTRW and GFPP on undirected networks} \label{CTRWnetworks}
\setcounter{section}{5} \setcounter{equation}{0} \setcounter{theorem}{0} 

\subsection{CTRW on undirected networks} 

In this section, we analyze random motions on undirected networks (graphs)
subordinated to a GFPP. In such a walk the jumps from one node to another take place at GFPP arrival times. To develop this model we employ
the CTRW approach by Montroll and Weiss \cite{MontrollWeiss1965} and consult also \cite{ScherLax1973}.
For general outlines and historical developments of the CTRW we refer to the references
\cite{KutnerMasoliver2017,Shlesinger2017}.
We first recall the Montroll-Weiss CTRW for undirected networks.
Then in Section \ref{definitionGFPP} we derive the generalization
of the fractional Kolmogorov-Feller equation for the walk subordinated to a GFPP. Finally, in Section \ref{genfractdiffusion} we explore this motion
on infinite $d$-dimensional integer lattices and analyze the `well-scaled diffusion limit'.

We consider an undirected connected network with $N$ nodes which we
denote with $p=1,\ldots ,N$. For our convenience, we employ Dirac's $|bra\rangle\langle ket|$ notation.
In an undirected network the positive-semidefinite $N \times N$
Laplacian matrix ${\mathbf L} =(L_{pq})$ characterizes
the network topology and is defined by \cite{TMM-APR-ISTE2019,NohRieger2004}
\vskip -13pt%
\begin{equation}
 \label{Laplacianmat}
 L_{pq}=K_p\delta_{pq}-A_{pq} ,
\end{equation}
\vskip -3pt \noindent
where ${\mathbf A} = (A_{pq})$ denotes the adjacency matrix having elements one
if a pair of nodes is connected and zero otherwise. Further we do not allow self-connections thus $A_{pp}=0$.
In an undirected network adjacency and Laplacian matrices are symmetric. The diagonal
elements $L_{pp}=K_p$ of the Laplacian matrix are referred to as the degrees
of the nodes $p$ counting the number of connected neighbor nodes with a node $p$ with $K_p=\sum_{q=1}^NA_{pq}$.
The one-step transition matrix ${\mathbf W} =(W_{pq})$ relating the
network topology with the probabilities for one jump in the random walk then is defined by \cite{TMM-APR-ISTE2019,NohRieger2004}
\vskip - 14pt %
\begin{equation}
 \label{one-step}
  W_{pq} =\frac{1}{K_p}A_{pq}= \delta_{pq}-\frac{1}{K_p}L_{pq},
\end{equation}
\vskip -3pt \noindent
which generally is a non-symmetric matrix for networks with variable degrees
$K_i\neq K_j$ $ (i\neq j)$.
The one-step transition matrix $W_{pq}$ defines the conditional probability that
the walker which is on node $p$ reaches in one step node $q$ \cite{TMM-APR-ISTE2019,NohRieger2004}.
We consider now a CTRW where the walker performs random steps (`jumps') governed by the one-step transition matrix $(W_{pq})$
at random arrival times $0< t_1<t_2< \ldots t_n< \ldots \infty$
in a renewal process (See Section \ref{CTRW}). We start the observation at $t=0$ where
walker sits on its departure node.

We introduce the $N\times N$ {\it transition matrix} ${\mathbf P}(t)= (P_{ij}(t))$
indicating the (dimensionless) {\it probability} to find the walker at time $t$ on node $j$
under the condition that the walker was sitting at node $i$ at time $t=0$ (start of the observation).
The transition matrix fulfills the (row-) normalization condition $\sum_{j=1}^NP_{ij}(t) =1$
and stochasticity implies that $0\leq P_{ij}(t) \leq 1$.
We restrict us on connected undirected networks and allow variable degrees.
In such networks the transition matrix is non-symmetric
$P_{ij}(t) \neq P_{ji}(t)$ (See \cite{TMM-APR-ISTE2019} for a detailed discussion).
Assuming an initial condition
${\mathbf P}(0)$, then the probability to find the walker
on node $j$ at time $t$ is given by the transition matrix $P_{ij}(t)$ which has the representation \cite{Cox1967}
\vskip - 12pt%
\begin{equation}
 \label{the-CTRW}
 {\mathbf P}(t) =  {\mathbf P}(0) \sum_{n=0}^{\infty} \Phi^{(n)}(t) {\mathbf W}^n,
\end{equation}
\vskip -3pt \noindent
where $\Phi^{(n)}(t)$ are the probabilities
that the walker within the time interval $[0,t]$ has made
$n$ steps (Eq. (\ref{nstepprobabilityuptot})) and ${\mathbf W}$ indicates the one-step transition matrix (\ref{one-step}).
The convergence of the series (\ref{the-CTRW}) can be easily proved by using that ${\mathbf W}$ has uniquely eigenvalues
$|\lambda_m|\leq 1$ \cite{TMM-APR-ISTE2019} together with the normalization of the state probabilities to see that
$|\sum_{n=0}^{\infty}\Phi^{(n)}(t)(\lambda_m)^n| \leq 1$.
We observe the relation
\vskip - 13pt%
\begin{align}
\nonumber
& {\mathbf P}(t) -\Phi^{0}(t){\mathbf P}(0)=
{\mathbf P}(0) \sum_{n=1}^{\infty}\Phi^{(n)}(t) {\mathbf W}^n \\ \nonumber
&= \int_0^t{\rm d}\tau  \chi(t-\tau) \left(\sum_{n=0}^{\infty}
 \Phi^{(n)}(\tau) {\mathbf P}(0) {\mathbf W}^n \right) {\mathbf W} \\
&=  \int_0^t{\rm d}\tau \chi(t-\tau) {\mathbf P}(\tau){\mathbf W}
 \label{andbyusing}
\end{align}
\vskip -4pt \noindent
with elements
\vskip - 13pt
\begin{equation*}
P_{ij}(t)=  \Phi^{0}(t) P_{ij}(0) + \int_{0}^t{\rm d}t' \chi(t-t')\sum_{\ell=1}^NP_{i\ell}(t')W_{\ell j} ,\hspace{0.5cm} t>0.
\end{equation*}
\vskip -2pt \noindent
This relation can be seen as the `CTRW on networks'- counterpart of the renewal equation (\ref{timedomainrel}) (For an outline of related aspects in renewal theory consult \cite{Cox1967,GorenfloMainardi2013,Gorenflo2010}).
We assume now the initial condition that the walker at $t=0$ sits on a known node $P_{ij}(0)=\delta_{ij}$. Then
the transition matrix (\ref{the-CTRW}) has the Laplace transform
\vskip -12pt%
\begin{equation}
 \label{occupationproblaplace}
 {\tilde {\mathbf P}}(s) = \frac{\left(1-{\tilde \chi}(s)\right)}{s} \left\{{\mathbf 1}-
 {\tilde \chi}(s){\mathbf W}\right\}^{-1}.
\end{equation}
\vskip -3pt \noindent%
It is useful to account for the canonical representation of the one-step
transition matrix \cite{TMM-APR-ISTE2019}
\vskip - 14pt%
\begin{equation}
 \label{canonic-one-step}
 {\mathbf W} = |\Phi_1\rangle\langle{\bar \Phi}_1|
 +\sum_{m=2}^N\lambda_m |\Phi_m\rangle\langle{\bar \Phi}_m| ,
\end{equation}
\vskip -3pt \noindent
where we have a unique eigenvalue $\lambda_1=1$ and
$|\lambda_m| < 1$ ($m=2,\ldots ,N$) (Consult \cite{TMM-APR-ISTE2019} for an analysis of the spectral structure).
In Eq. (\ref{canonic-one-step}) we have introduced the right- and
left eigenvectors $|\Phi_j\rangle$ and $ \langle{\bar \Phi}_j|$ of ${\mathbf W}$,
respectively. The first term $|\Phi_1\rangle\langle{\bar \Phi}_1|$ corresponding to $\lambda_1=1$
is the stationary distribution.
Hence the canonical representation of (\ref{occupationproblaplace}) has the form
\vskip -10pt %
\begin{equation}
 \label{canonic-occupation-probabilities}
 \ds {\tilde {\mathbf P}}(s)=
 \frac{\left(1-{\tilde \chi}(s)\right)}{s}\sum_{m=1}^N
 \frac{|\Phi_m\rangle\langle{\bar \Phi}_m|}{(1- \lambda_m{\tilde \chi}(s))}
\end{equation}
\vskip -4pt \noindent
with the eigenvalues
\vskip - 10pt%
\begin{equation}
\label{montroll-eiss-eigenvalues}
 {\tilde P}(m,s)=
 \frac{1}{s} \frac{\left(1-{\tilde \chi}(s)\right)}{(1- \lambda_m \,{\tilde \chi}(s))} ,\hspace{0.8cm} m=1,\ldots, N.
\end{equation}
We mention that for $\lambda_1=1$ we have ${\tilde P}(1,s) = s^{-1}$
thus in the time domain the stationary amplitude yields $P(1,t)=\mathcal{L}^{-1}(s^{-1}) =\Theta(t)$.
Equation (\ref{montroll-eiss-eigenvalues})
is the celebrated
{\it Montroll-Weiss formula} \cite{MontrollWeiss1965,ScherLax1973,Shlesinger2017}
and is of utmost importance in a wide range of fields.

\subsection{Generalized fractional Kolmogorov-Feller equation} 
\label{definitionGFPP}
In this section, we specify the renewal process in the CTRW to a GFPP and derive
a generalization of the Laskin's fractional Kolmogorov-Feller equation
\cite{Laskin2003} describing the time evolution of the transition matrix.
The GFPD probabilities $\Phi^{(n)}_{\beta,\alpha}(t)$ have with Eq. (\ref{jump-gen-fractional-laplacetr})
the Laplace transforms
\vskip -10pt %
\begin{equation}
 \label{alplacephin}
 {\tilde \Phi}^{(n)}_{\beta,\alpha}(s) = {\tilde \Phi}^{(0)}_{\beta,\alpha}(s)\frac{\xi^{n\alpha}}{(s^{\beta}+\xi)^{n\alpha}} , \hspace{0.5cm} \alpha >0 ,\hspace{0.5cm} 0< \beta\leq 1.
\end{equation}
For $n=0$ we have the Laplace transform of the survival probability
\begin{equation}
 \label{Phizeroalhbet}
 {\tilde \Phi}^{(0)}_{\beta,\alpha}(s) =\frac{1-{\tilde \chi}_{\beta,\alpha}(s)}{s}   =
 \frac{(s^{\beta}+\xi)^{\alpha}-\xi^{\alpha}}{s(s^{\beta}+\xi)^{\alpha}} .
\end{equation}
These Laplace transforms fulfill the following relations
\vskip -10pt
\begin{equation}
 \label{generalrelationngen}
 (s^{\beta}+\xi)^{\alpha} {\tilde \Phi}^{(n)}_{\beta,\alpha}(s) = \xi^{\alpha} {\tilde \Phi}^{(n-1)}_{\beta,\alpha}(s) ,\hspace{1cm} n=1,2,\ldots ,
\end{equation}
and for $n=0$ we have
\vskip -10pt %
\begin{equation}
 \label{Phizer-oalhbetnz-zero-case}
 (s^{\beta}+\xi)^{\alpha} {\tilde \Phi}^{(0)}_{\beta,\alpha}(s) = \frac{(s^{\beta}+\xi)^{\alpha}-\xi^{\alpha}}{s} .
\end{equation}
Now the goal is to derive the generalized counterpart of the fractional
Kolmogorov-Feller equation which is obtained with the causal time domain representations of relations
(\ref{generalrelationngen}) and (\ref{Phizer-oalhbetnz-zero-case}).
To this end we introduce the convolution kernel
\vskip -10pt %
\begin{equation}
 \label{kernel-associated}
\mathcal{D}^{\beta,\alpha}(t) = \mathcal{L}^{-1} \left\{(s^{\beta}+\xi)^{\alpha}\right\} .
\end{equation}
In the causal time domain Eq. (\ref{generalrelationngen}) takes the form
\begin{equation}
\label{takeforms}
_0\!\mathcal{D}_t^{\beta,\alpha} \cdot \Phi^{(n)}_{\beta,\alpha}(t) = \xi^{\alpha} \Phi^{(n-1)}_{\beta,\alpha}(t) ,\hspace{0.8cm} n=1,2,\ldots ,
\end{equation}
\vskip -3pt \noindent
where
$_0\!\mathcal{D}_t^{\beta,\alpha} \cdot \Phi(t)=:
\int_0^t\Phi(\tau)\mathcal{D}^{\beta,\alpha}(t-\tau){\rm d}\tau$
has to be read as a convolution. The time domain representation of this kernel
is obtained explicitly as (See Appendix \ref{derivation-convolutionGFPP} for a derivation, especially
Eqs. (\ref{representbyfracderiv})-(\ref{we-then-get}))
\vskip - 11pt %
\begin{align}
\nonumber
& \mathcal{D}^{\beta,\alpha}(t) = \mathcal{L}^{-1} \left\{(s^{\beta}+\xi)^{\alpha}\right\}
= \frac{d^{\ceil{\alpha\beta}}}{dt^{\ceil{\alpha\beta}}}\left\{\Theta(t)  {\mathcal B}_{\beta,\alpha}(t)\right\}
\\
&= \frac{d^{\ceil{\alpha\beta}}}{dt^{\ceil{\alpha\beta}}}
 \left\{ \Theta(t) t^{\ceil{\alpha\beta}-\beta\alpha-1}
 E_{\alpha,\beta,(\ceil{\alpha\beta}-\alpha\beta)}(\xi t^{\beta}) \right\}
 \hspace{0.5cm}\mathrm{for}\hspace{0.4cm} \alpha\beta \notin \mathbb{N} ,
 \label{en-res-mittag-leffl-gn-kernel-explicit1}
\end{align}
\vskip -3pt \noindent
and
\vskip - 13pt %
\begin{equation}
 \label{en-res-mittag-leffl-gn-kernel-explicit2}
 \mathcal{D}^{\beta,\alpha}(t) =\frac{d^{\alpha\beta}}{dt^{\alpha\beta}}\left(\delta(t) +
\Theta(t)\frac{d}{dt}E_{\alpha,\beta,1}(\xi t^{\beta})\right)
\hspace{0.5cm}\mathrm{for}\hspace{0.2cm}  \alpha\beta \in \mathbb{N},
\end{equation}
\vskip -3pt \noindent
where $E_{a,b,c}(z)$ is a Prabhakar Mittag-Leffler type function defined in Eq.
(\ref{Mittag-Leffler-other-gen}). We emphasize that the $\Theta(t)$-function must
be taken `under the time derivative' of integer order $\ceil{\alpha\beta}$ where the ceiling function $\ceil{\gamma}$ indicates the smallest integer greater or equal to $\gamma$.
Above distributional representations for $\mathcal{D}^{\beta,\alpha}(t)$ define then convolution operators in the sense of relation (\ref{convkern}) and related properties of Laplace transforms are
outlined in Appendix \ref{AppendLaplacetrafo} (See property (\ref{it-follows-that}) together with decomposition
(\ref{pointdep})).
For $n=0$ from Eq. (\ref{Phizer-oalhbetnz-zero-case}) we get
\vskip - 13pt %
\begin{equation}
 \label{nzeroyields}
 _0\!\mathcal{D}_t^{\beta,\alpha}(\xi) \Phi^{(0)}_{\beta,\alpha}(t) =
 \mathcal{L}^{-1}\left\{\frac{(s^{\beta}+\xi)^{\alpha}-\xi^{\alpha}}{s}\right\} .
\end{equation}
\vskip -4pt \noindent
The kernel
\vskip - 13pt %
\begin{equation}
 \label{K0kernelalphabeta}
\mathcal{L}^{-1}\left\{\frac{(s^{\beta}+\xi)^{\alpha}-\xi^{\alpha}}{s}\right\} =
K^{(0)}_{\beta,\alpha}(t)
 -\xi^{\alpha}\Theta(t)
\end{equation}
\vskip -3pt \noindent
is evaluated explicitly in Appendix \ref{further-kernel} (Eq. (\ref{K0CaseB}))
where $K^{(0)}_{\beta,\alpha}(t)$ is obtained as
\vskip - 13pt %
\begin{align}
\nonumber
&K^{(0)}_{\beta,\alpha}(t) =\mathcal{L}^{-1}\left\{\frac{(s^{\beta}+\xi)^{\alpha}}{s}\right\} =
\frac{d^{\ceil{\alpha\beta}-1}}{dt^{\ceil{\alpha\beta}-1}}\left\{\Theta(t)  {\mathcal B}_{\beta,\alpha}(t)\right\}
\\
&= {\left\{\begin{array}{l} \Theta(t)t^{-\alpha\beta}
 E_{\alpha,\beta,1-\alpha\beta}(\xi t^{\beta}) , \hspace{1cm} 0<\alpha\beta <1 \\[4pt]
 \frac{d^{\ceil{\alpha\beta}-1}}{dt^{\ceil{\alpha\beta}-1}}
   \left\{\Theta(t) t^{\ceil{\alpha\beta}-\beta\alpha-1}
 E_{\alpha,\beta,(\ceil{\alpha\beta}-\alpha\beta)}(\xi t^{\beta})\right\}
 \hspace{0.3cm} \alpha\beta >1, \hspace{0.1cm} \alpha\beta \notin \mathbb{N}
 \\ [4pt] 
 \frac{d^{\alpha\beta-1}}{dt^{\alpha\beta-1}}\left(\delta(t) +
\Theta(t)\frac{d}{dt}E_{\alpha,\beta,1}(\xi t^{\beta})\right),
\hspace{1cm} \alpha\beta  \in \mathbb{N}
                                    \end{array}\right.  }
 \label{Kzeroexplicitform}
\end{align}
\vskip -4pt \noindent
and we have $\mathcal{D}^{\beta,\alpha}(t)=\frac{d}{dt}K^{(0)}_{\beta,\alpha}(t)$. In all these relations the expressions holding for $\alpha\beta \in \mathbb{N}$ are recovered by the expressions for $\alpha\beta\notin \mathbb{N}$ in the limit $\alpha\beta \rightarrow \ceil{\alpha\beta}-0$.
These kernels contain the Prabhakar type function
\begin{equation}
 \label{Mittag-Leffler-other-gen}
 \begin{array}{l}
 \ds E_{c,a,b}(z) = E_{a,b}^{-c}(-z) = \sum_{m=0}^{\infty} \frac{(-c)_m}{m!} \frac{(-z)^m}{\Gamma(a b+b)} \\
 \hspace{0.1cm}
 \ds=
 \sum_{m=0}^{\infty} \frac{c!}{(c-m)!m!} \frac{z^m}{\Gamma(a b+b)}  ,\hspace{0.5cm}
 \Re\{a\} >0, \hspace{0.1cm} \Re\{b\} >0  ,\hspace{0.1cm} a, b, c, z \, \in \mathbb{C},
 \end{array}
\end{equation}
where $E_{u,v}^w(z)$ is defined in (\ref{genmittag-Leff}).
The transition probability matrix (\ref{the-CTRW})
of the GFPP walk then writes
\vskip - 11pt %
\begin{equation}
\label{the-CTRW-GFPP}
 {\mathbf P}_{\beta,\alpha}(t) = {\mathbf P}(0) \Phi_{\beta,\alpha}^{(0)}(t)   +
 {\mathbf P}(0) \sum_{n=1}^{\infty} \Phi_{\beta,\alpha}^{(n)}(t) {\mathbf W}^n .
\end{equation}
\vskip -3pt \noindent
Then we obtain with (\ref{takeforms}), (\ref{nzeroyields}) the equation
\begin{align}
\nonumber
 _0\!\mathcal{D}_t^{\beta,\alpha} \cdot {\mathbf P}_{\beta,\alpha}(t) &= {\mathbf P}(0)  \mathcal{L}^{-1}\left\{\frac{(s^{\beta}+\xi)^{\alpha}-\xi^{\alpha}}{s}\right\} +\xi^{\alpha} {\mathbf P}(0) \sum_{n=1}^{\infty} \Phi_{\beta,\alpha}^{(n-1)}(t) {\mathbf W}^n \\ &=
{\mathbf P}(0)  \mathcal{L}^{-1}\left\{\frac{(s^{\beta}+\xi)^{\alpha}-\xi^{\alpha}}{s}\right\} +
\xi^{\alpha} {\mathbf P}_{\beta,\alpha}(t){\mathbf W}
\label{generalizedfracKolmogorivFeller}
\end{align}
which takes in the causal time domain the form ($0<\beta \leq 1$, $\alpha >0$)
\begin{equation}
 \label{explict-gen-fract}
 _0\!\mathcal{D}_t^{\beta,\alpha} \cdot {\mathbf P}_{\beta,\alpha}(t) = {\mathbf P}(0) \left(K^{(0)}_{\beta,\alpha}(t)
 -\xi^{\alpha}\Theta(t)\right)  + \xi^{\alpha} {\mathbf P}_{\beta,\alpha}(t){\mathbf W}.
 \end{equation}
 We call equation (\ref{explict-gen-fract}) `{\it generalized fractional Kolmogorov-Feller equation}'.
 This equation can be rewritten as
 \begin{align}
 \nonumber
   &\frac{d^{\ceil{\alpha\beta}-1}}{dt^{\ceil{\alpha\beta}-1} }  \left\{\frac{d}{dt}\int_0^t
  {\mathcal B}_{\beta,\alpha}(\tau){\mathbf P}_{\beta,\alpha}(t-\tau){\rm d\tau} -
  {\mathcal B}_{\beta,\alpha}(t){\mathbf P}(0) \right\} + \xi^{\alpha}{\mathbf P}(0)  \\[1mm]
  &= \xi^{\alpha} {\mathbf P}_{\beta,\alpha}(t){\mathbf W} , \hspace{1cm} t\in \mathbb{R}_{+}
   \label{rewriteGFKFeqs}
 \end{align}
 to give
 \begin{equation}
 \label{GKFresult}
    \frac{d^{\ceil{\alpha\beta}-1}}{dt^{\ceil{\alpha\beta}-1} }  \int_0^t
  {\mathcal B}_{\beta,\alpha}(t-\tau)\frac{d}{d\tau}{\mathbf P}_{\beta,\alpha}(\tau){\rm d}\tau +
  \xi^{\alpha} {\mathbf P}(0) = {\mathbf P}_{\beta,\alpha}(t){\mathbf W}.
 \end{equation}
 This representation of the generalized fractional Kolmogorov-Feller equation
 emphasizes the connection with Prabhakar fractional calculus \cite{dosSantos2019,Giusti2020,Giusti-et-al2020,PolitoTomovski2016}.
 Taking into account the explicit representations
(\ref{en-res-mittag-leffl-gn-kernel-explicit1}) and (\ref{Kzeroexplicitform}) this equation writes
for $\alpha\beta \notin {\mathbb{N}}$
 \begin{align}
\nonumber
  &\frac{d^{\ceil{\alpha\beta}}}{dt^{\ceil{\alpha\beta}}} \int_0^t
 (t-\tau)^{\ceil{\alpha\beta}-\beta\alpha-1}
 E_{\alpha,\beta,(\ceil{\alpha\beta}-\alpha\beta)}(\xi (t-\tau)^{\beta})
 {\mathbf P}_{\beta,\alpha}(\tau){\rm d}\tau\\
  &= {\mathbf P}(0) \left(K^{(0)}_{\beta,\alpha}(t)
 -\xi^{\alpha}\Theta(t)\right)  +
 \xi^{\alpha} {\mathbf P}_{\beta,\alpha}(t){\mathbf W}  .
 \label{explicitGKFequation}
\end{align}
\medskip 
{\sl Laskin fractional Poisson process limit $\alpha=1$}:

Let us consider (\ref{explict-gen-fract}) with (\ref{explicitGKFequation}) for the Laskin fractional case $\alpha=1$ and $0<\beta<1$.
Then we have (See (\ref{Kzeroexplicitform}))
\vskip -11pt%
\begin{equation}
 \label{kernel0alphabet}
 K^{(0)}_{\beta,1}(t) = {\mathcal B}_{\beta,1}(t) =  t^{-\beta}E_{1,\beta,1-\beta}(\xi t^{\beta})  =  \frac{t^{-\beta}}{\Gamma(1-\beta)} + \xi ,\hspace{0.5cm} t>0 ,
\end{equation}
\vskip -2pt \noindent
thus Eq. (\ref{GKFresult}) takes the form
\begin{equation}
 \label{KFfractcasealpha1-caputo}
\frac{1}{\Gamma(1-\beta)}  \int_0^t(t-\tau)^{-\beta}
 \frac{d}{d\tau}{\mathbf P}_{\beta,1}(\tau){\rm d}\tau
= -\xi {\mathbf P}_{\beta,1}(t) + \xi {\mathbf P}_{\beta,1}(t){\mathbf W}.
\end{equation}
The left-hand side of this equation is the Caputo-fractional derivative of order $\beta$.
This equation can also be written as (corresponding to (\ref{explicitGKFequation}))
\begin{equation}
 \label{fractKGFeq}
 _0 \!D_t^{\beta}{\mathbf P}_{\beta,1}(t) - \frac{t^{-\beta}}{\Gamma(1-\beta)} {\mathbf P}(0) = -\xi {\mathbf P}_{\beta,1}(t) +
 \xi {\mathbf P}_{\beta,1}(t){\mathbf W},
\end{equation}
where $_0\!D_t^{\beta}$ is the Riemann-Liouville fractional derivative of order $\beta$.
(\ref{KFfractcasealpha1-caputo}) and (\ref{fractKGFeq}) are equivalent representations of the {\it fractional Kolmogorov-Feller equation} derived by Laskin and others \cite{BeghinOrsinger2009,GorenfloMainardi2013,Gorenflo2010,HilferAnton1995,Laskin2003,MainardiGorenfloScalas2004,MetzlerKlafter2000,SaichevZaslavski1997,Zaslavsky2002}.
By accounting for $\lim_{\beta\rightarrow 1-0}\frac{t^{-\beta}}{\Gamma(1-\beta)}=\delta(t)$
in (\ref{KFfractcasealpha1-caputo})) one can see that the fractional K-F equation for $\alpha=1$ and $\beta=1$
reduces to the classical K-F equation for the walk subordinated to the standard Poisson process.
In this way we have shown that the generalized fractional K-F equation contains for $\alpha=1$ the well-known classical counterparts.
%
%
\section{Generalized fractional diffusion in the infinite \break $d$-dimensional integer lattice}
\label{genfractdiffusion}

\setcounter{section}{6} \setcounter{equation}{0} \setcounter{theorem}{0} 

In this section, we analyze the effects of the GFPP on diffusive features
in the infinite $d$-dimensional integer lattice ${\mathbb{Z}}^d$.
Especially we derive here the `well-scaled' diffusion limit.
We denote with $\vec{q}=(q_1,..,q_d)$ and $q_j=0, \pm 1, \pm 2, \ldots \in {\mathbb{Z}_0} $ the lattice points (nodes of the network).
The canonical representation of the one-step transition
matrix (\ref{canonic-one-step})
becomes then a Fourier integral
over the $d$-dimensional Brillouin zone $k_j \in [-\pi,\pi]$
and has the eigenvalues \cite{TMM-APR-ISTE2019,TMM-APR-JPhys-A2017}
\vskip - 11pt%
\begin{equation}
 \label{eigvals}
 \begin{array}{l}
\ds  \lambda(\vec{k})= 1-\frac{1}{2d}\mu(\vec{k}) ,\hspace{1cm} \mu(\vec{k}) = 2d-2\sum_{j=1}^d\cos(k_j), \\
\ds  \lambda(\vec{k}) = \frac{1}{d}\sum_j^d\cos(k_j) ,\hspace{1cm} -\pi \leq k_j \leq \pi ,
 \end{array}
\end{equation}
\vskip -3pt \noindent
where $\mu(\vec{k})$ indicates the eigenvalues of the Laplacian matrix (\ref{Laplacianmat})
of the lattice. We assumed that any node is connected only with its $2d$ closest neighbor nodes.
The Laplace transform (\ref{canonic-occupation-probabilities}) of the transition matrix then writes
\vskip - 12pt%
\begin{equation}
 \label{greensfunction}
 {\mathbf P}(s,\vec{q}) =
 \frac{\left(1-{\tilde \chi}_{\beta,\alpha}(s)\right)}{(2\pi)^d s}
 \int_{-\pi}^{\pi}{\rm d}k_1\ldots \int_{-\pi}^{\pi}{\rm d}k_d
 \frac{e^{i\vec{k}\cdot\vec{q}}}{(1- \lambda(\vec{k}){\tilde \chi}_{\beta,\alpha}(s))} .
\end{equation}
We assume here the initial condition that the walk starts in the origin $\vec{q}=\vec{0}$.
We subordinate a normal random walk on the infinite $d$-dimensional integer lattice
to a GFPP.
In order to analyze diffusive motions we account for the behavior of
the eigenvalues (\ref{eigvals})
for $k \rightarrow 0$ (where $k=|\vec{k}|$), namely
\vskip -10pt%
\begin{equation}
 \label{transitionlaceigs}
 \lambda(\vec{k}) \approx 1-\frac{1}{2d}k^2 .
\end{equation}
Then we get for the Montroll-Weiss
equation (\ref{montroll-eiss-eigenvalues}) for $k\rightarrow 0$ small
\vskip - 12pt%

\begin{equation}
 \label{PKS}
 \begin{array}{l}
\ds {\tilde P}(k,s) = \frac{\left(1-{\tilde \chi}_{\beta,\alpha}(s)\right)}{s}\frac{1}{(1-{\tilde \chi}_{\beta,\alpha}(s)\lambda(k))} =
 \frac{s^{-1}}{1+\frac{{\tilde \chi}_{\beta,\alpha}(s)}{(1-{\tilde \chi}_{\beta,\alpha}(s))}\frac{k^2}{2d}} \\ \\
\ds  \hspace{0.5cm} \approx \left(\frac{1}{s}-
 \frac{{\tilde \chi}_{\beta,\alpha}(s)}{s (1-{\tilde \chi}_{\beta,\alpha}(s))}\frac{k^2}{2d} +\ldots \right) =
 \frac{1}{s} -{\tilde {\bar n}}_{\beta,\alpha}(s) \frac{k^2}{2d} +\ldots .
\end{array}
 \end{equation}
This relation contains the Laplace transform ${\tilde {\bar n}}_{\beta,\alpha}(s)$ of the expected number of arrivals (\ref{Laplacetranfostepsaverge}).
Before we further exploit this relation let us first derive the diffusion equation which is emerging in a `well-scaled diffusion limit'.
To this end we rewrite the exact Montroll-Weiss equation (\ref{PKS}) as follows
\vskip -12pt %
\begin{multline}
 \label{diff-relation1}
  -\frac{\xi^{\alpha}}{2d} \mu(\vec{k}) {\tilde P}_{\beta,\alpha}(k,s) = (s^{\beta}+\xi)^{\alpha} \left({\tilde P}_{\beta,\alpha}(k,s)-{\tilde \Phi}^{(0)}_{\beta,\alpha}(s)\right) \\
-\xi^{\alpha}{\tilde P}_{\beta,\alpha}(k,s)  .
\end{multline}
We notice that the matrix representation of this equation gives a generalized fractional K-F equation
of the type (\ref{explict-gen-fract}).
Eq. (\ref{diff-relation1}) takes for small $k$ the representation
\vskip - 12pt %
\begin{multline}
 \label{diff-relation2}
 - \frac{h^2}{2d} {\bar k}^2   {\tilde P}_{\beta,\alpha}(h{\bar k},s) \approx \left((1+\frac{1}{\xi}s^{\beta})^{\alpha}-1\right){\tilde P}_{\beta,\alpha}(h{\bar k},s) \\
 +\frac{1}{s}\left(1-(1+\frac{1}{\xi}s^{\beta})^{\alpha}\right),
\end{multline}
\vskip -2pt \noindent
where we utilized that the Laplacian eigenvalues $\mu(\vec{k}) \approx k^2$ for $k$ small (See (\ref{eigvals})).
Eq. (\ref{diff-relation2}) holds asymptotically for small (dimensionless wave numbers) $k={\bar k}h$ where we introduce
a new wave vector ${\vec {\bar k}} =({\bar k}_1,..,{\bar k}_d)$ ($k_j \in \mathbb{R}$ having physical dimension $[cm]^{-1}$).
Now we consider the `well-scaled diffusion limit' of
(\ref{diff-relation2}). Generally there are different ways (not necessarily well-scaled procedures) to define the diffusion limit leading to different types of diffusion equations
\cite{Gorenflo2010,MichelitschRiascosPhysA2020,MichelitschRiascos-etal2020,TMM-APR-ISTE2019}.
Let us introduce the Laplace operator $\Delta = \sum_{j=1}^d \frac{\partial^2}{\partial x_j^2}$ of the $\mathbb{R}^d$
with respect to the new rescaled quasi-continuous spatial coordinates
$\vec{x} =h\vec{p} \in \mathbb{R}^d$ (having units $[cm]$). Then we introduce $P_{(\beta,\alpha)}(\vec{q},t) = h^{d} \mathcal{P}_{(\beta,\alpha)}(\vec{x},t)$ where $\mathcal{P}_{(\beta,\alpha)}(\vec{x},t)$
indicates the spatial transition probability kernel having physical units $[cm]^{-d}$.
For any finite ${\bar k}$ we can choose $h>0$ sufficiently small such that
$\mu(\vec{k}) \approx k^2= ({\bar k}h)^2 \rightarrow 0$ thus
the left-hand side of relation (\ref{diff-relation2}) tends to zero for $h\rightarrow 0$. To maintain the right-hand side in relation (\ref{diff-relation2}) also small requires that $\xi(h) \rightarrow \infty$
for $h\rightarrow 0$ thus we can expand in equation (\ref{diff-relation2}) $(1+\frac{1}{\xi}s^{\beta})^{\alpha} \approx 1+ \frac{\alpha}{\xi}s^{\beta}$ leading to relation
\vskip -10pt%
\begin{equation}
\label{lim}
 - \frac{h^2\xi }{2d\alpha} {\bar k}^2   {\tilde  P}_{\beta,\alpha}(h{\bar k},s) \approx s^{\beta}
 {\tilde P}_{\beta,\alpha}(h{\bar k},s) - s^{\beta-1} .
\end{equation}
In order the left-hand side of (\ref{lim}) remains finite for $h\rightarrow 0$ requires the scaling
$\xi(h) \sim \xi_0 h^{-2} \rightarrow \infty$ (where $\xi_0>0$ is a new constant independent of $h$).

The constant $\frac{\xi_0}{2d\alpha}$ has physical dimension
$[cm^2sec^{-\beta}]$ and can be interpreted as a generalized fractional diffusion constant (recovering for $\beta=1$ the units $[cm^2sec^{-1}]$ of normal diffusion).
We observe that the index $\alpha$ enters Eq. (\ref{lim}) only as a scaling parameter.
The well-scaled diffusion limiting equation (\ref{lim}) then
leads to the fractional diffusion equation
\vskip -10pt%
\begin{equation}
\label{genfracdiffeq}
  \frac{\xi_0}{2d\alpha}\Delta \mathcal{P}_{\beta,\alpha}(\vec{x},t)  = _0 \!D_t^{\beta} \cdot \mathcal{P}_{\beta,\alpha}(\vec{x},t) -
  \delta^{(d)}(\vec{x}) \,\frac{t^{-\beta}}{\Gamma(1-\beta)},
  \end{equation}
\vskip -1pt \noindent
where $\mathcal{P}_{\beta,\alpha}(\vec{x},t)= \delta^{(d)}(\vec{x})$ is the initial condition in the form of a Dirac's $\delta$-function in $\mathbb{R}^d$ and $\vec{x} =h\vec{q} \in \mathbb{R}^d$ are the
rescaled quasi-continuous coordinates. $ _0 \!D_t^{\beta}\cdot  P_{\beta,\alpha}(q,t)$ indicates the Riemann-Liouville fractional derivative of order $\beta$ (See Appendix \ref{derivation-convolutionGFPP}).
The right-hand side of (\ref{genfracdiffeq}) can be read as
Caputo-fractional derivative of order $\beta$ of $\mathcal{P}_{\beta,\alpha}(\vec{x},t)$ where
the term $\,\frac{t^{-\beta}}{\Gamma(1-\beta)}$ reflects the non-Markovian feature
with long-time memory of the walk (For a discussion consult \cite{MichelitschRiascosPhysA2020,MichelitschRiascos-etal2020,MichelitschPolitoRiascos2020}).

We notice that the well-scaled diffusive limit of
generalized fractional diffusion leads to the same type of purely fractional diffusion equation
(\ref{genfracdiffeq})
as for the fractional Poisson process ($\alpha=1$ with $0<\beta<1$) having Mittag-Leffler waiting times. (\ref{genfracdiffeq}) hence reflects `asymptotic Mittag-Leffler universality' \cite{GorenfloMainardi2006}.
Indeed Eq. (\ref{genfracdiffeq}) coincides
with
the fractional diffusion equation widely used in fractional dynamics
\cite{MetzlerKlafter2004,Zaslavsky2002} (and many others).

Finally we consider the standard Poisson limit $\alpha=1$ and $\beta=1$. Then we get with (\ref{lim})
Fick's second law of normal diffusion
\begin{equation}
 \label{Fick2}
 \frac{\xi_0}{2d}\Delta \mathcal{P}_{1,1}(\vec{x},t)  = \frac{\partial }{\partial t} \mathcal{P}_{1,1}(\vec{x},t) - \delta(t) \delta^{(d)}(\vec{x}),
\end{equation}
where $\frac{\xi_0}{2d}$ indicates the diffusion constant of standard (normal-) diffusion.
This equation is also obtained by the limit $\beta\rightarrow 1-0$ in Eq (\ref{genfracdiffeq})
taking into account $\lim_{\beta\rightarrow 1-0}\frac{t^{-\beta}}{\Gamma(1-\beta)} = \delta(t)$.
This behavior reflects the Markovian memoryless nature of underlying the pure Poisson process.

Finally let us consider the mean squared displacement in generalized fractional diffusion.
Then we obtain the Laplace-Fourier transform of the mean squared displacement ${\tilde \sigma^2}(s)$
by taking into account (\ref{PKS}) for $\xi(h) \sim \xi_0h^{-2}$ large
\vskip -10pt%
\begin{equation}
 \label{meandisplacementsGFPP}
 {\tilde \sigma^2}_{\beta,\alpha}(s) = -\Delta_{{\vec {\bar k}}} {\tilde P}(h{\bar k},s)\Big|_{{\bar k}=0} = h^2 {\tilde {\bar n}}_{\beta,\alpha}(s,h) \sim \frac{\xi_0}{\alpha}s^{-\beta-1},
\end{equation}
where $\Delta_{{\vec {\bar k}}}$ stands for the Laplace operator with respect to the new wave vector $\vec{{\bar k}} \in \mathbb{R}^d$.
For the GFPP the asymptotic expression for the expected number of arrivals
${\bar n}_{\beta,\alpha}(t)$ was determined in
(\ref{largetimesteps}). We hence obtain for the mean square displacement
\begin{equation}
 \label{tworegimesGFPP}
\ds \sigma^2_{\beta,\alpha}(t)
= \frac{ \xi_0t^{\beta} }{\alpha \Gamma(\beta+1)}   ,
\hspace{0.5cm} 0 < \beta \leq 1 \hspace{0.5cm} \alpha >0.
\end{equation}
We have for $0<\beta<1$ and for all $\alpha>0$ including fractional Poisson $\alpha=1$
sublinear $t^{\beta}$ power-law behavior
corresponding to fat-tailed jump density (\ref{tlarge-fat-tailed}).
The sublinear power law $\sigma_{\beta,\alpha}^2(t) \sim t^{\beta}$ ($0<\beta<1$) corresponds also to the universal limit for $t\xi^{\frac{1}{\beta}} \rightarrow\infty$ of fat-tailed waiting time PDFs where extremely long waiting times occur.
Such behavior is well known in the literature for anomalous {\it subdiffusion}
($0<\beta<1$) \cite{MetzlerKlafter2000}.

In contrast for $\beta=1$
for all $\alpha >0$ (Erlang regime which includes the standard
Poisson $\alpha=1$, $\beta=1$) we obtain normal diffusive behavior with linear
increase $\sigma^2_{\beta,\alpha}(t) =
\frac{\xi_0}{\alpha} t$ of the mean squared displacement.
This behavior can be attributed to {\it normal diffusion} \cite{MetzlerKlafter2000}.
The emergence of universal scaling laws for $t\xi^{\frac{1}{\beta}}\rightarrow \infty$ indeed
reflects the asymptotic Mittag-Leffler universality for time
fractional dynamics \cite{GorenfloMainardi2006}.
%
%
\section{Conclusions}

We have analyzed a generalization of the Laskin fractional Poisson process, the
{\it generalized fractional Poisson process (GFPP)} and derived the
{\it generalized fractional Poisson distribution (GFPD)} (Eq.(\ref{Generalized-Fractional-Poisson-Distribution})). The GFPP is (unless in the standard Poisson case  $\alpha=1$ and $\beta=1$) non-Markovian and introduces long-time memory effects.
For $\alpha=1$, $0<\beta<1$ the fractional Poisson process, for $\beta=1$, $\alpha>0$ the Erlang process,
and for $\alpha=1$, $\beta=1$ the (memoryless) standard Poisson process are recovered.
We showed that for $\alpha=1$ the GFPD state probabilities reduce to Laskin's fractional
Poisson distribution (\ref{fractionalpoission-distribution}) and for $\alpha=1$ with $\beta=1$ to the standard Poisson distribution.
 The GFPP contains two index parameters $\alpha >0$ and $0< \beta \leq 1$ and a characteristic time scale $\xi^{-\frac{1}{\beta}}$
 controlling the dynamic behavior. For long dimensionless observation times the
asymptotic universal scaling of the fractional Poisson process emerges in the range $0<\beta<1$, $\alpha>0$
(asymptotic Mittag-Leffler universality).

We developed the Montroll-Weiss CTRW with GFPP (Prabhakar distributed) waiting times between the jumps and analyzed the resulting stochastic motions in
undirected networks. We derived the `{\it generalized fractional
Kolmogorov-Feller equation}' (Eqs. (\ref{explict-gen-fract})-(\ref{GKFresult}))
which governs the time-evolution of the transition matrix in undirected networks.

As an application we analyzed for the $d$-dimensional infinite integer lattice the `well-scaled diffusion limit' and the resulting stochastic motion. We obtained the same type of fractional diffusion equation as for CTRWs with Mittag-Leffler distributed waiting times in a fractional Poisson process.
All equations turn in the limits $\alpha=1$, $0<\beta<1$,
and $\beta=1$, $\alpha=1$ into their classical counterparts
of Laskin's fractional Poisson process and standard Poisson process, respectively.

The GFPP has a huge potential of further applications in anomalous diffusion and transport phenomena including turbulence,
non-Markovian stochastic motions, and in the dynamics of complex systems.

\vspace*{-5pt}

\appendix
\renewcommand{\theequation}{\Alph{section}.\arabic{equation}} 
 %
 

\section{Laplace transforms of causal functions} \label{AppendLaplacetrafo}
\setcounter{equation}{0} 

In this appendix, we discuss some properties related with
causal distributions and generalized functions \cite{GelfangShilov1968}
and their Laplace transforms. First we introduce the Heaviside step function
\vskip -11pt %
\begin{equation}
 \label{Haviside}
 \Theta(t) = \left\{ \begin{array}{l} 1 ,\hspace{1cm} t \geq 0, \\[2mm]
                      0 ,\hspace{1cm} t<0,
                     \end{array}\right.
\end{equation}
\vskip -3pt \noindent
where we emphasize that $\Theta(0)=1$ and we have $\delta(t)= \frac{d}{dt}\Theta(t)$ for Dirac's $\delta$-distribution.
We consider uniquely {\it causal} functions and distributions $f(t)=\Theta(t)f(t)$ which
may take non-zero values only for non-negative times $t\geq 0$.
First of all we introduce the Laplace transform ${\tilde f}(s)$ of a causal function $f(t)=\Theta(t)f(t)$
by
\begin{equation}
 \label{laplacetranfocausalfunction}
 {\tilde f}(s) = \mathcal{L}\{f(t)\}= \int_{0-}^{\infty} e^{-st} \Theta(t) f(t){\rm d}t
 ,\hspace{0.5cm} s=\sigma+i\omega,
\end{equation}
with $\sigma=\Re\{s\} > \sigma_0$. We choose the lower integration limit infinitesimally negative $0-$ thus $\Theta(0-)= \lim_{\epsilon\rightarrow 0+}\Theta(-\epsilon)=0$. We denote by $\mathcal{L}^{-1}\left\{{\tilde f}(s)\right\}=f(t)$ the inverse Laplace transform.
As an important issue in our analysis let us compare the two Laplace transforms (i)
$\mathcal{L}\{\frac{d^m}{dt^m}\left[\Theta(t)f(t)\right]\}$
and (ii) $\mathcal{L}\{\Theta(t)\frac{d^m}{dt^m}f(t)\}$  where  $m\in \mathbb{N}_0$ is integer or zero.
One obtains straightforwardly (i) by $m$ partial integrations
\begin{equation}
 \label{it-follows-that}
 \ds \mathcal{L}\left\{\frac{d^m}{dt^m}\left[\Theta(t)f(t)\right]\right\} =  s^m {\tilde f}(s).
\end{equation}
On the other hand, we have for (ii)
\vskip - 11pt %
\begin{equation}
 \label{case-ii}
  \mathcal{L}\left\{\Theta(t)\frac{d^m}{dt^m}f(t)\right\} =
  s^m {\tilde f}(s)-\sum_{k=0}^{m-1} s^{m-1-k} \frac{d^k}{dt^k}f(t)\Big|_{t=0} .
\end{equation}
In the analysis of this paper we often suppress writing the Heaviside $\Theta(t)$-function in causal functions
in cases where no derivatives are involved.
Further we mention that we utilize both notations $\Gamma(\beta)=(\beta-1)!$ for the $\Gamma$-function.
\vspace*{-3pt}
\section{Some properties of the fractional Poisson process} \label{fractionalPoissonprocess}
\setcounter{equation}{0} 

Now we are interested in the state probabilities (\ref{nstepprobabilityuptot}), i.e. the probabilities for $n$ arrivals within interval $[0,t]$
for the fractional Poisson process with Mittag-Leffler density (\ref{Mittag-Leffler-waiting-time-PDF}).
This probability is given by
\vskip -10pt%
\begin{equation}
 \label{thisproba}
 \Phi^{(n)}_{\beta}(t)= \int_0^{t}E_{\beta}(-\xi(t-\tau)^{\beta}) \chi^{(n)}_{\beta}(\tau){\rm d}\tau
\end{equation}
\vskip -3pt \noindent
where the Laplace transform is with ${\tilde \chi}_{\beta}^{(n)}(s)= \chi_{\beta}^n(s)=\frac{\xi^n}{(\xi+s^{\beta})^n}$,
\vskip -10pt%
\begin{equation}
 \label{Laplacetrafoofofnsteps}
 {\tilde \Phi}^{(n)}_{\beta}(s) = \frac{1-{\tilde \chi}_{\beta}(s)}{s}{\tilde \chi}_{\beta}^n(s) =\frac{\xi^ns^{\beta-1}}{(\xi + s^{\beta})^{n+1}} .
\end{equation}
\vskip -2pt \noindent
For evaluation it is convenient to employ the generating function of relation (\ref{generatingsteps})
\vskip - 12pt%
\begin{equation}
 \label{genfuMittag-Leffler}
 G_{\beta}(t,v)= \sum_{n=0}^{\infty} v^n \Phi^{(n)}_{\beta}(t) ,\hspace{1cm} |v| \leq 1,
\end{equation}
\vskip -3pt \noindent
and its Laplace transform
\begin{align}
\nonumber
 {\tilde G}_{\beta}(s,v)&= \sum_{n=0}^{\infty} v^n {\tilde \Phi}^{(n)}_{\beta}(s) = \frac{(1-{\tilde \chi}_{\beta}(s))}{s}
 \frac{1}{(1-v{\tilde \chi}_{\beta}(s))} \\
&= \frac{s^{\beta-1}}{\xi(1-v)+s^{\beta}},
\label{genfuLaplaceMittagLeffler}
\end{align}
\vskip -2pt \noindent
where (\ref{Laplacetrafoofofnsteps}) can be expressed by $\frac{1}{n!}\frac{d^n}{dv^n}{\tilde G}_{\beta}(s,v){\big|}_{v=0}$.
Again from ${\tilde G}_{\beta}(s,1)= s^{-1}$ follows that the $\Phi^{(n)}_{\beta}(t)$ are a normalized probability distribution on $n=\{0,1,2,\ldots \infty\}$, i.e. $G_{\beta}(t,v){\Big|}_{v=1}= \sum_{n=0}^{\infty} \Phi^{(n)}_{\beta}(t) = 1$.
It follows from (\ref{survival}) that Laplace inversion of (\ref{genfuLaplaceMittagLeffler})
yields again a Mittag-Leffler function where we have to replace in $\xi \rightarrow \xi(1-v)$ in
(\ref{survival}). In this way we obtain for the generating function (\ref{genfuMittag-Leffler})
the Mittag-Leffler function
\begin{equation}
\label{Mittag-Leffler-gen-function}
 G_{\beta}(t,v) =   E_{\beta}(-\xi(1-v)t^{\beta})
 \hspace{0.5cm} t\geq 0 ,\hspace{0.5cm} 0<\beta \leq 1 .
\end{equation}
This expression was also obtained by Laskin \cite{Laskin2003}.
The probabilities $\Phi^{(n)}_{\beta}(t)$ are then obtained by
\begin{equation}
 \label{nsteps}
  \Phi^{(n)}_{\beta}(t) =\frac{1}{n!} \frac{{\rm d}^n}{{\rm d}v^n}G_{\beta}(t,v){\Big|}_{v=0}=
  \frac{1}{n!} \frac{{\rm d}^n}{{\rm d}v^n}E_{\beta}(\xi(v-1)t^{\beta}){\Big|}_{v=0}  .
\end{equation}
This relation yields the {\it fractional Poisson distribution},
i.e. the probability for $n$ arrivals within $[0,t]$ in a fractional Poisson
process ($0<\beta \leq 1$) \cite{Laskin2003}
\begin{equation}
 \label{fractionalpoission}
 \Phi^{(n)}_{\beta}(t)   =
 \frac{(\xi t^{\beta})^n}{n!} \sum_{m=0}^{\infty}\frac{(m+n)!}{m!}\frac{(-\xi t^{\beta})^m}{(\beta(m+n))!} ,\hspace{0.5cm} n\in \mathbb{N}_0,
\end{equation}
\vskip -2pt \noindent
where $n=0$ gives the Mittag-Leffler survival probability $\Phi^{(0)}(t)= E_{\beta}(-\xi t^{\beta})$.
For $\beta=1$ relation (\ref{fractionalpoission}) recovers the {\it standard Poisson distribution}
\begin{equation}
\label{standardpoisson}
\Phi^{(n)}_{\beta=1}(t) =  \frac{(\xi t)^n}{n!} e^{-\xi t}
\end{equation}%
\vskip -3pt \noindent
with the generating function (\ref{Mittag-Leffler-gen-function})
\vskip -12pt%
\begin{equation}
 \label{genfubeta1}
 G_{\beta=1}(t,v) =  e^{-\xi t} \sum_{n=0}^{\infty} \frac{(\xi t)^n}{n!} = e^{(v-1)\xi t}.
\end{equation}
The normalization of the fractional Poisson distribution is easily verified by
\vskip - 13pt%
\begin{equation}
 \label{normalmittag}
 \sum_{n=0}^{\infty} \Phi^{(n)}_{\beta}(t) = G_{\beta}(t,v){\Big|}_{v=1} = E_{\beta}((v-1)\xi t^{\beta})_{v=1} = 1.
\end{equation}
\vskip -3pt \noindent
The expected number of arrivals (See Eq. (\ref{averageNumberSteps})) in a
fractional Poisson process within time interval $[0,t]$ is obtained as
\begin{align}
\nonumber
 {\bar n}_{\beta} (t) &= \sum_{n=0}^{\infty} n \Phi^{(n)}_{\beta}(t) = \frac{{\rm d}}{{\rm d}v}G_{\beta}(t,v){\Big|}_{v=1} =
 \frac{{\rm d}}{{\rm d}v} E_{\beta}((v-1) \xi t^{\beta}){\Big|}_{v=1} \\
&= \frac{\xi t^{\beta}}{\Gamma(\beta+1)}.
 \label{averagefractionalsteps}
\end{align}
%
%

\section{Evaluation of the kernel $\mathcal{L}^{-1} \left\{(s^{\beta}+\xi)^{\alpha}\right\} $}
\label{derivation-convolutionGFPP}
\setcounter{equation}{0} 

Let us make a brief general remark on notations. Convolution kernels that we write in the distributional representation
$k(t)=\frac{d^m}{dt^m}\left[\Theta(t)f(t)\right]$, $m \in \mathbb{N}_0$ (having Laplace transform ${\tilde k}(s)=s^m{\tilde f}(s)$), where the Heaviside step function is included into the differentiation
define convolutions as follows
\begin{align}
    \nonumber
    \int_0^tk(t-\tau)g(\tau) {\rm d}\tau =
    \int_{-\infty}^{\infty} \Theta(t-\tau) g(t-\tau) \frac{d^m}{d\tau^m}\left[\Theta(\tau)f(\tau)\right] {\rm d}\tau  \\
     = \frac{d^m}{dt^m} \int_0^{t} f(t-\tau)g(\tau){\rm d}\tau ,\hspace{0.8cm} t \in \mathbb{R}_{+}.
     \label{convkern}
    \end{align}
On the other hand if $k(t)$ appears as function and is not integrated we can read it in on $\mathbb{R}_{+}$
as $k(t)=\frac{d^m}{dt^m}\left[\Theta(t)f(t)\right]= \frac{d^m}{dt^m}f(t)$.
\smallskip 

Now let us evaluate the time domain representation of the generalized fractional operator of Eq. (\ref{kernel-associated})
defined by
\begin{equation}
 \label{kernel-associated-def}
\mathcal{D}^{\beta,\alpha}(t) =  \mathcal{L}^{-1} \left\{(s^{\beta}+\xi)^{\alpha}\right\} ,\hspace{0.5cm} \alpha>0,
\hspace{0.5cm} 0<\beta \leq 1, \hspace{0.5cm} \xi >0 .
\end{equation}
The following simple representation will turn out to be highly useful to evaluate kernel (\ref{kernel-associated-def}), namely
\begin{equation}
 \label{pointdep}
(s^{\beta}+\xi)^{\alpha} = s^{\beta\alpha}(1+\xi s^{-\beta})^{\alpha} =
s^{m} s^{\alpha\beta -m}(1+\xi s^{-\beta})^{\alpha}
\end{equation}
where in the evaluation the important property (\ref{it-follows-that}) will be employed.
We have to consider two possible cases: (i) $\alpha\beta$ is not integer $\alpha\beta \notin \mathbb{N} $ and
(ii) $\alpha\beta$ is integer $\alpha\beta \in \mathbb{N} $.

Let us first evaluate case (i)  $\alpha\beta \notin \mathbb{N} $.
To this end it is convenient to introduce the {\it ceiling} function $\ceil{(..)}$ where $m= \ceil{\lambda}$
indicates the
smallest integer larger or equal to $\lambda$. In this way by putting $m=\ceil{\alpha\beta}$ in relation
(\ref{pointdep}) we see that in the
part $s^{\alpha\beta -\ceil{\alpha\beta}}(1+\xi s^{-\beta})^{\alpha}$ of the decomposition
occur terms $s^{-\gamma}$ with negative powers $-\gamma<0$
leading to inverse Laplace transforms $\mathcal{L}^{-1}\{s^{-\gamma}\}=
\Theta(t) \frac{t^{\gamma-1}}{\Gamma(\gamma)}$ giving (up to coefficients)
Riemann-Liouville fractional integral kernels where
$\gamma- 1 \geq \ceil{\alpha\beta}-\alpha\beta-1 > -1 $, i.e. the terms are either weakly singular
or non-singular and hence in both cases integrable on $\mathbb{R}_{+}$.
With these observations let us now
derive the causal time domain representation
of the kernel (\ref{kernel-associated-def}).
We can write
\begin{align}
\nonumber
&\mathcal{D}^{\beta,\alpha}(t)= \mathcal{L}^{-1}\left\{s^{\ceil{\alpha\beta}}
 s^{\alpha\beta-\ceil{\alpha\beta}}(1+\xi s^{-\beta})^{\alpha}\right\}\\ \nonumber
&=\frac{d^{\ceil{\alpha\beta}}}{dt^{\ceil{\alpha\beta}}} \mathcal{L}^{-1}\left\{
 s^{\alpha\beta-\ceil{\alpha\beta}}(1+\xi s^{-\beta})^{\alpha}\right\}, \hspace{0.5cm} s=\sigma+i\omega,
 \hspace{0.2cm} \sigma >0,
\\
& = e^{\sigma t}\left(\sigma+\frac{d}{dt}\right)^{\ceil{\alpha\beta}}(\sigma+\frac{d}{dt})^{\alpha\beta-\ceil{\alpha\beta}}
 \left(1+(\sigma+\frac{d}{dt})^{-\beta}\xi\right)^{\alpha}\delta(t)
 \label{rewritethegenfracop}
\end{align}
taking the form of a Fourier integral ($m=\ceil{\alpha\beta}$)
\vskip -10pt%
\begin{multline}
 \label{Fouriergenfracop}
 \mathcal{D}^{\beta,\alpha}(t) =
 e^{\sigma t}\left(\sigma+\frac{d}{dt}\right)^m \\ \times
\int_{-\infty}^{\infty}\frac{{\rm d}\omega}{(2\pi)}e^{i\omega t}
 (\sigma+i\omega)^{\alpha\beta-m} \left(1+(\sigma+i\omega)^{-\beta}\xi\right)^{\alpha}.
\end{multline}
Then for $\sigma > \xi^{\frac{1}{\beta}}$ we can expand (\ref{Fouriergenfracop}) with respect to
$(\sigma+i\omega)^{-\beta}\xi$ and integrate each term to arrive at
\begin{equation}
\label{en-res-mittag-leffl-gn-kernel}
 \begin{array}{l}
  \mathcal{D}^{\beta,\alpha}(t) = \mathcal{L}^{-1} \left\{(s^{\beta}+\xi)^{\alpha}\right\}  \\ \\  
  =
  \frac{d^{\ceil{\alpha\beta}}}{dt^{\ceil{\alpha\beta}}} \left\{ \Theta(t) t^{\ceil{\alpha\beta}-\beta\alpha-1} \sum_{n=0}^{\infty}\frac{\alpha!}{(\alpha-n)!n!}
 \frac{(\xi  t^{\beta})^n}{\Gamma(\beta n + \ceil{\alpha\beta} -\beta\alpha)} \right\} \\ \\
 = \left\{ \Theta(t) t^{\ceil{\alpha\beta}-\beta\alpha-1} E_{\alpha,\beta,(\ceil{\alpha\beta}-\alpha\beta)}(\xi t^{\beta}) \right\} ,\hspace{0.2cm} \alpha\beta \notin \mathbb{N}, 
 \end{array}
\end{equation}
where we notice that $-1< {\ceil{\alpha\beta}}-\beta\alpha-1 <0$.
This relation contains the Prabhakar
type function (\ref{Mittag-Leffler-other-gen})
which converges in the entire $z$-plane.
Let us consider order $n=0$ in (\ref{en-res-mittag-leffl-gn-kernel}) which can be evaluated in the same way as above
($\gamma=\alpha\beta$) where $\gamma >0$ with $\gamma \notin \mathbb{N}$, namely
\begin{equation}
 \label{part1}
 \begin{array}{l}
 \mathcal{L}^{-1}\{s^{\gamma}\}=  \mathcal{L}^{-1} \{s^{\ceil{\gamma}} s^{\gamma -\ceil{\gamma}}\}
= e^{\sigma t}\left(\sigma+\frac{d}{dt}\right)^{\ceil{\gamma}}\left(\sigma+\frac{d}{dt}\right)^{\gamma-\ceil{\gamma}} \delta(t)
   \\  \\
= e^{\sigma t}\left(\sigma+\frac{d}{dt}\right)^{\ceil{\gamma}} \left\{ e^{-\sigma t}\Theta(t)
 \frac{t^{\ceil{\gamma}-\gamma-1}}{(\ceil{\gamma}-\gamma -1)!}\right\}  =
 \frac{d^{\ceil{\gamma}}}{dt^{\ceil{\gamma}}} \left(\Theta(t)
 \frac{t^{\ceil{\gamma}-\gamma-1}}{\Gamma(\ceil{\gamma}-\gamma)} \right).
 \end{array}
\end{equation}
We identify this expression with the kernel of the Riemann-Liouville fractional
derivative of order $\gamma$  (e.g. \cite{MillerRoss1993} and many others) which defines a convolution
$_0\! D^{\gamma}_t\cdot f(t) $ in the sense of (\ref{convkern}), namely
\begin{equation}
 \label{fractionalderivative}
 _0\! D^{\gamma}_tf(t)= \frac{1}{\Gamma(\ceil{\gamma}-\gamma)}
 \frac{d^{\ceil{\gamma}}}{dt^{\ceil{\gamma}}} \int_0^{t}(t-\tau)^{\ceil{\gamma}-\gamma-1}f(\tau){\rm d}\tau .
\end{equation}
One observes that integer-order derivatives are also covered
\begin{equation}
\label{R-L-fracderlim}
\lim_{\gamma\rightarrow \ceil{\gamma}-0}\, _0\! D^{\gamma}_t f(t)=
\frac{d^{\ceil{\gamma}}}{dt^{\ceil{\gamma}}}f(t)
\end{equation}
where (\ref{part1}) then takes $\lim_{\gamma\rightarrow \ceil{\gamma}-0}\frac{d^{\ceil{\gamma}}}{dt^{\ceil{\gamma}}}\delta(t)$ ($\ceil{\gamma}=\gamma$).
\smallskip 

Then to complete our demonstration let us finally evaluate also the part
\begin{equation}
 \label{thepart2}
 \begin{array}{l}
   \mathcal{L}^{-1}\{ (1+\xi s^{-\beta})^{\alpha}\}   =
 e^{\sigma t} \left(1+(\sigma+\frac{d}{dt})^{-\beta}\xi\right)^{\alpha}\delta(t)
  \\  
  =
 \delta(t)+  \Theta(t) \frac{d}{dt}\sum_{n=0}^{\infty}\frac{\alpha !}{(\alpha-n)!n!}
 \frac{\xi^n t^{n\beta}}{\Gamma(n\beta+1)} \\ 
 =
 \delta(t) +\Theta(t) \frac{d}{dt}E_{\alpha,\beta,1}(\xi t^{\beta}) = \frac{d}{dt} \left[
\Theta(t)E_{\alpha,\beta,1}(\xi t^{\beta})\right].
 \end{array}
\end{equation}
We notice that in this series the order $n=0$
yields Dirac's $\delta$-distribution $\delta(t)$ and $E_{\alpha,\beta,1}(\xi t^{\beta})$ is
a Prabhakar type function defined
in Eq. (\ref{Mittag-Leffler-other-gen}).
From above considerations (See Eq. (\ref{pointdep})) it follows that we can represent the kernel
(\ref{en-res-mittag-leffl-gn-kernel}) by the fractional Riemann-Liouville derivative of order $\alpha\beta$ as
\vskip -10pt%
\begin{align}
\nonumber
 &  \small  \mathcal{D}^{\beta,\alpha}(t) =  _0\! D^{\alpha\beta}_t \{ \delta(t)+\Theta(t)\frac{d}{dt}E_{\alpha,\beta,1}(\xi t^{\beta}) \} \hspace{0.5cm}  \alpha\beta \notin \mathbb{N}   \\ \nonumber
&
 \small = \frac{1}{\Gamma(\ceil{\alpha\beta}-\alpha\beta)}
 \frac{d^{\ceil{\alpha\beta}}}{dt^{\ceil{\alpha\beta}}}
 \int_0^{t}(t-\tau)^{\ceil{\alpha\beta}-\alpha\beta-1}\left( \delta(\tau) +\frac{d}{d\tau}E_{\alpha,\beta,1}(\xi \tau^{\beta}) \right) {\rm d}\tau
 \\ \nonumber
&
 \small = \frac{d^{\ceil{\alpha\beta}}}{dt^{\ceil{\alpha\beta}}} \left\{ \Theta(t)\frac{t^{\ceil{\alpha\beta}-\alpha\beta-1}}{\Gamma(\ceil{\alpha\beta}-\alpha\beta)}
 + \Theta(t) \sum_{n=1}^{\infty} \frac{\alpha !}{(\alpha-n)!n!} \frac{\xi^n t^{\beta n +\ceil{\alpha\beta}-\alpha\beta-1}}{\Gamma(\beta n+\ceil{\alpha\beta}-\alpha\beta)} \right\}
 \\ \nonumber
&
 \small = \frac{d^{\ceil{\alpha\beta}}}{dt^{\ceil{\alpha\beta}}}
 \left\{\Theta(t) t^{\ceil{\alpha\beta}-\beta\alpha-1} E_{\alpha,\beta,(\ceil{\alpha\beta}-\alpha\beta)}(\xi t^{\beta}) \right\}\\
& = \frac{d^{\ceil{\alpha\beta}}}{dt^{\ceil{\alpha\beta}}}\left\{ \Theta(t){\mathcal B}_{\beta,\alpha}(t)\right\},
 \label{representbyfracderiv}
\end{align}
where
\vskip - 13pt%
\begin{align}
\label{prabhakarkernel}
 {\mathcal B}_{\beta,\alpha}(t) &=
 \mathcal{L}^{-1}\left\{
 s^{\alpha\beta-\ceil{\alpha\beta}}(1+\xi s^{-\beta})^{\alpha}\right\} \\[1mm]  \nonumber
& =  t^{\ceil{\alpha\beta}-\beta\alpha-1}
 E_{\alpha,\beta,(\ceil{\alpha\beta}-\alpha\beta)}(\xi t^{\beta}) \qquad \alpha\beta \notin \mathbb{N}.
\end{align}
We notice that ${\mathcal B}_{\beta,\alpha}(t)$ is a (weakly singular) `{\it Prabhakar kernel}'
[We refer to the recent paper of Giusti \cite{Giusti2020} for an introduction of `Prabhakar generalized fractional calculus'] where
${\mathcal B}_{\beta,\alpha}(t)=e^{-\alpha}_{\beta,\ceil{\alpha\beta}-\alpha\beta}(-\xi,t)$ (in his notation).
\smallskip 

Now let us evaluate case (ii) where $ \ceil{\alpha\beta}= \alpha\beta \in \mathbb{N} $ is integer. We obtain then
\vskip -11pt %
\begin{align}
\nonumber
 & {\mathcal B}_{\beta,\alpha}(t) =
 \mathcal{L}^{-1}\left\{
 s^{\alpha\beta-\ceil{\alpha\beta}}(1+\xi s^{-\beta})^{\alpha}\right\} = \mathcal{L}^{-1}
 \left\{(1+\xi s^{-\beta})^{\alpha}\right\} \\
 &= \delta(t) +
\Theta(t)\frac{d}{dt}E_{\alpha,\beta,1}(\xi t^{\beta})
 = \frac{d}{dt} \left(
\Theta(t)E_{\alpha,\beta,1}(\xi t^{\beta})\right), \hspace{2mm}\alpha\beta \in \mathbb{N}.
  \label{integercaseA}
\end{align}
Since for $\alpha\beta \in \mathbb{N}$ we have $\ceil{\alpha\beta}=\alpha\beta$ thus this expression coincides with (\ref{thepart2}). Then we get
\begin{align}
\nonumber
\mathcal{D}^{\beta,\alpha}(t) &= \frac{d^{\ceil{\alpha\beta}}}{dt^{\ceil{\alpha\beta}}}    \left( \Theta(t){\mathcal B}_{\beta,\alpha}(t) \right) \\
& =
\frac{d^{\ceil{\alpha\beta}}}{dt^{\ceil{\alpha\beta}}}
\left(\delta(t) +
\Theta(t)\frac{d}{dt}E_{\alpha,\beta,1}(\xi t^{\beta})\right),
\hspace{0.5cm} \alpha\beta \in \mathbb{N} .
\label{we-then-get}
\end{align}
This result is consistent with (\ref{representbyfracderiv}) and is recovered from this expression in the limiting case
$\alpha\beta \rightarrow \ceil{\alpha\beta}-0$ when we account for the limit
of the
fractional Riemann-Liouville derivative (\ref{R-L-fracderlim}) reducing then to an integer order derivative.
%
%

\section{Evaluation of
$\mathcal{L}^{-1}\left\{\frac{(s^{\beta}+\xi)^{\alpha}-\xi^{\alpha}}{s}\right\}$}
\label{further-kernel}
\setcounter{equation}{0} 
Here we evaluate the kernel
\begin{equation}
 \label{nzeroyields22}
 \mathcal{L}^{-1}\left\{\frac{(s^{\beta}+\xi)^{\alpha}-\xi^{\alpha}}{s}\right\} =
 K^{0}_{\beta,\alpha}(t)
 -\xi^{\alpha}\Theta(t), 
\end{equation}
that occurs in (\ref{nzeroyields}) which is now straightforward accounting
for the results in previous Appendix \ref{derivation-convolutionGFPP}.
Hence we arrive at
\vskip -10pt
 \begin{equation}
  \label{K0CaseB}
  \begin{array}{l}
  \ds K^{0}_{\beta,\alpha}(t) = \mathcal{L}^{-1}\left\{\frac{(s^{\beta}+\xi)^{\alpha}}{s}\right\} =
 \frac{d^{\ceil{\alpha\beta}-1}}{dt^{\ceil{\alpha\beta}-1}} \left(\Theta(t)
 {\mathcal B}_{\beta,\alpha}(t)\right) \\ \\  \hspace{0.5cm}
 =\left\{\begin{array}{l}
    =   \frac{d^{\ceil{\alpha\beta}-1}}{dt^{\ceil{\alpha\beta}-1}}  \left(\Theta(t)
    t^{\ceil{\alpha\beta}-\beta\alpha-1}
 E_{\alpha,\beta,(\ceil{\alpha\beta}-\alpha\beta)}(\xi t^{\beta})\right), \hspace{0.5cm} \alpha\beta \notin \mathbb{N}, \\ \\  =  \frac{d^{\ceil{\alpha\beta}-1}}{dt^{\ceil{\alpha\beta}-1}} \left(\delta(t) +
\Theta(t)\frac{d}{dt}E_{\alpha,\beta,1}(\xi t^{\beta})\right),  \hspace{0.5cm} \alpha\beta  \in \mathbb{N} . \end{array}\right.
 \end{array}
 \end{equation}
For $\alpha\beta <1 $ we have $\ceil{\alpha\beta}-1=0$ thus in that case
$K^{0}_{\beta,\alpha}(t)= {\mathcal B}_{\beta,\alpha}(t)$.
We observe in these relations that $\mathcal{D}^{\beta,\alpha}(t)=\frac{d}{dt}K^{0}_{\beta,\alpha}(t)$.
%



 %

%


\bigskip
\smallskip

 \it
 \noindent
$^{1, \S}$ Sorbonne Universit\'e \\ Institut Jean le Rond d'Alembert, CNRS UMR 7190 \\
4 place Jussieu, 75252 Paris cedex 05, FRANCE \\[4pt]
e-mail: michel@lmm.jussieu.fr (Corresp. author)  \hfill  Received: June 24, 2019\\
\hspace*{1cm} \hfill Revised:  June 17, 2020\\ 
 \\
$^2$ Instituto de F\'isica \\ 
Universidad Nacional Aut\'onoma de M\'exico \\
Apartado Postal 20-364, 01000 Ciudad de M\'exico, M\'EXICO \\[4pt]
e-mail: aperezr@fisica.unam.mx


\begin{thebibliography}{99} \normalsize 

\bibitem{Angstmann-et-al2017} 
C. N. Angstmann, B. I. Henry, B. A. Jacobs, A. V. McGann,
A time-fractional generalised advection equation from a stochastic process. 
\emph{Chaos, Solitons $\&$ Fractals} \textbf{102} (2017), 175--183.

\bibitem{BarkaiCheng2003} E. Barkai, Y.-C. Cheng, Aging continuous time random walks. 
\emph{J. Chem. Phys.} \textbf{118} (2003), Art. \# 6167.

\bibitem{BarkaiMetzlerKlafter2000} E. Barkai, R. Metzler, and J. Klafter, From continuous time
random walks to the fractional Fokker-Planck equation. 
\emph{Phys. Rev. E} \textbf{61}, No 1 (2000), Art. \# 132.

\bibitem{BeghinOrsinger2009} L. Beghin, E. Orsingher, Fractional
Poisson processes and related random motions. 
\emph{Electron. J. Probab.} \textbf{14} (2009), Art. \# 61, 1790--1826.


\bibitem{Brungelson1989} J.D. Bryngelson, P. G. Wolynes,
Intermediates and barrier crossing in a random energy model (with applications to protein folding). 
\emph{J. Chem. Phys.} \textbf{93} (1989), Art. \# 19, 6902--6915.

\bibitem{CahoyPolito2013} D.O. Cahoy, F. Polito, Renewal processes based on generalized Mittag-Leffler waiting times. \emph{Commun. Nonlinear Sci. Numer. Simul.} \textbf{18}, No 3 (2013), 639--650.

\bibitem{ChechkinHofmannSokolov2009} A.V. Chechkin, M. Hoffmann, I.M. Sokolov, Continuous-time random walk with correlated waiting times. \emph{Phys. Rev. E} \textbf{80} (2009),
Art. \# 031112.

\bibitem{Cox1967} D.R. Cox, \emph{Renewal Theory}, Methuen. London (1967); ISBN 041220570X.

\bibitem{dosSantos2019} M.A. dos Santos, Fractional Prabhakar Derivative in Diffusion Equation with Non-Static Stochastic Resetting. \emph{Physics} \textbf{1}, No 1 (2019), 40--58.

\bibitem{Feller1971}
W. Feller, \emph{An Introduction to Probability Theory and Its Applications} \textbf{1}, John Wiley \& Sons, New York (1968); ISBN: 978-0-471-25708-0.

\bibitem{GarraGarrappa2018} R. Garra, R. Garrappa, The Prabhakar or three parameter Mittag-Leffler function: Theory and application. \emph{Commun. Nonlinear Sci. Numer. Simul.} \textbf{56} (2018), 314--329.

\bibitem{GarraGorenfloPolito2014} R. Garra, R. Gorenflo, F. Polito, Z. Tomovski, Hilfer-Prabhakar derivatives and some applications.\emph{ Appl. Math. Comput.} \textbf{242} (2014), 576--589.

\bibitem{GelfangShilov1968} I. Gel'fand, G.E. Shilov, \emph{Generalized Functions I-III}, Academic Press, New York, (1968); ISBN 1-4704-2659-5.

\bibitem{Giusti2020} A. Giusti, General fractional calculus and Prabhakar’s theory.
\emph{Commun. Nonlinear Sci. Numer. Simul.} \textbf{83} (2020), Art. \# 105114.

\bibitem{Giusti-et-al2020} A. Giusti, I. Colombaro, R. Garra, R. Garrappa, F. Polito, M. Popolizio, F. Mainardi, A practical guide to Prabhakar fractional calculus. \emph{Fract. Calc. Appl. Anal.} \textbf{23}, No 1 (2020), 9--54.

\bibitem{GiustiColombaro2018} A. Giusti, I. Colombaro, Prabhakar-like fractional viscoelasticity. \emph{Commun. Nonlinear Sci. Numer. Simul.} \textbf{56} (2018), 138--143.

\bibitem{GorenfloKilbas2014} R. Gorenflo, A.A. Kilbas, F. Mainardi, S.V. Rogosin, \emph{Mittag-Leffler Functions, Related Topics and Applications}. Springer, New York (2014); ISBN: 978-3-662-43929-6.

\bibitem{GorenfloMainardi2013} R. Gorenflo, F. Mainardi, On the Fractional Poisson Process and the Discretized Stable Subordinator. \emph{Axioms} \textbf{4}, No 3 (2015), 321--344.

\bibitem{Gorenflo2010} R. Gorenflo, Mittag-Leffler Waiting Time, Power Laws, Rarefaction, Continuous Time Random Walk, Diffusion Limit. \emph{arXiv:1004.4413} (2010).

\bibitem{GorenfloMainardiVivoli2007} R. Gorenflo, F. Mainardi, A. Vivoli, Continuous time random walk and parametric subordination in fractional diffusion. \emph{Chaos, Solitons $\&$ Fractals}, \textbf{34} (2007), 89--103.

\bibitem{GorenfloMainardi2006} R. Gorenflo, F. Mainardi, The asymptotic universality of the Mittag-Leffler waiting time law in continuous time random walks, Invited lecture. 373. WE-Heraeus-Seminar on Anomalous Transport, Bad-Honnef (Germany), 12--16 July 2006.

\bibitem{Gorenflo2007} R. Gorenflo, E. A.A. Abdel Rehim, From Power Laws to Fractional Diffusion: the Direct Way. \emph{Vietnam J. Math.}, \textbf{32} (2004), 65--75.

\bibitem{HauboldMathaiSaxena2011} H.J. Haubold, A.M. Mathhai, R.K. Saxena, Mittag-Leffler functions and their applications.\emph{J. Appl. Math.} (2011), Art. \# 298628.

\bibitem{HilferAnton1995} R. Hilfer, L. Anton, Fractional master equations and fractal time random walks. \emph{Phys. Rev. E} \textbf{51}, No 2 (1995), R848--R851.

\bibitem{KilbasSrivastavaTruillo2006} A.A. Kilbas, H.M. Srivastava and J.J. Trujillo,
\emph{Theory and Applications of Fractional Differential Equations}, Elsevier, Amsterdam (2006); ISBN 9780444518323.

\bibitem{KlagesRadonsSokolov2008}  R. Klages, G. Radons, I.M. Sokolov,
\emph{Anomalous Transport: Foundations and Applications}. Wiley-VCH, Weinheim (2008); 
ISBN 9783527622986. %

\bibitem{ZumofenShlesingerKlafter1996} J. Klafter, M. F. Shlesinger and G. Zumofen, Beyond Brownian Motion. \emph{Phys. Today} \textbf{49}, No 2 (1996) 33--39.

\bibitem{KutnerMasoliver2017} R. Kutner, J. Masoliver, The continuous time random walk, still trendy: fifty-year history, state of art and outlook. \emph{Eur. Phys. J. B} \textbf{90} (2017), Art. \# 50.

\bibitem{Laskin2003} N. Laskin, Fractional Poisson process. \emph{Commun. Nonlinear Sci. Numer. Simul.} \textbf{8} (2003), 201--213.

\bibitem{Laskin2009} N. Laskin, Some applications of the fractional Poisson probability distribution. \emph{J. Math. Phys.} \textbf{50} (2009), Art. \# 113513.

\bibitem{MainardiGarrappa2015}
F. Mainardi, R. Garrappa, On complete monotonicity of the Prabhakar function and non-Debye relaxation in dielectrics. \emph{J. Comput. Phys.} \textbf{293} (2015), 70--80.

\bibitem{MainardiGorenfloScalas2004} F. Mainardi, R. Gorenflo, E. Scalas, A fractional generalization of the Poisson processes. \emph{Vietnam J. Math.} 
\textbf{32} (2004), 53--64.

\bibitem{Mathai2010} A. M. Mathai, Some properties of Mittag-Leffler functions and matrix variant analogues: A statistical perspective. \emph{Fract. Calc. Appl. Anal.} \textbf{13}, No. 2 (2010), 113--132.

\bibitem{MeerschaertEtal2011} M.M. Meerschaert, E. Nane, P.Villaisamy, The Fractional Poisson Process and the Inverse Stable Subordinator. \emph{Electron. J. Probab.} \textbf{16} (2011),
Art. \# 59.

\bibitem{MetzlerKlafter2004} R. Metzler, J. Klafter, The restaurant at the end of the random walk: recent developments in the description of anomalous
transport by fractional dynamics. \emph{J. Phys. A: Math. Gen.} \textbf{37} (2004), R161--R208.

\bibitem{MetzlerKlafter2000} R. Metzler, J. Klafter, The Random Walk's Guide to Anomalous Diffusion: A Fractional Dynamics Approach. \emph{Phys. Rep.} \textbf{339} (2000), 1--77.

\bibitem{MichelitschRiascosPhysA2020} T.M. Michelitsch, A.P Riascos, Continuous time random walk and diffusion with generalized fractional Poisson process. \emph{Physica A}
\textbf{545} (2020), Art. \# 123294.

\bibitem{MichelitschRiascos-etal2020} T.M. Michelitsch, A.P. Riascos, B.A. Collet, A.F. Nowakowski, F.C.G.A. Nicolleau, Generalized space-time fractional dynamics in networks and lattices. In: \emph{Nonlinear Wave Dynamics of Materials and Structures. Advanced Structured Materials} \textbf{122} Springer, Cham (2020); ISBN 978-3-030-38707-5.

\bibitem{MichelitschPolitoRiascos2020} T.M. Michelitsch, F. Polito, A.P. Riascos, On discrete-time generalized fractional Poisson process and related stochastic dynamics. \emph{arXiv:2005.06925} (2020).

\bibitem{TMM-APR-ISTE2019} T. Michelitsch, A. P\'erez Riascos, A. Nowakowski, F. Nicolleau, 
\emph{Fractional Dynamics on Networks and Lattices}, ISTE-Wiley (2019); ISBN 9781786301581.

\bibitem{TMM-APR-JPhys-A2017} T.M. Michelitsch, B.A. Collet, A.P. Riascos,
A.F. Nowakowski, F.C.G.A. Nicolleau, On recurrence of random walks with long-range steps generated by fractional Laplacian matrices on regular networks and simple cubic lattices.
\emph{J. Phys. A Math. Theor.} \textbf{50}, No 50 (2017), Art. \# 505004.

\bibitem{MillerRoss1993} K. S. Miller, B. Ross. \emph{An Introduction to the Fractional Calculus and Fractional Differential Equations.}, John Wiley $\&$ Sons, New York (1993); 
ISBN: 978-0471588849.

\bibitem{MonthusBouchaud1996} C. Monthus, J. P. Bouchaud, Models of traps and glass phenomenology. \emph{J. Phys. A Math. Gen.} \textbf{29}, No 14 (1996), 3847--3869.

\bibitem{MontrollWeiss1965} W.W. Montroll, G.H. Weiss, Random walks on lattices II.
\emph{J. Math. Phys.} \textbf{6}, No 2 (1965), 167--181.

\bibitem{NohRieger2004} J.D. Noh, H. Rieger, Random walks on complex networks. 
\emph{Phys. Rev. Lett.} \textbf{92}, No. 11. (2004), Art. \# 118701.

\bibitem{OldhamSpanier1974} K.B. Oldham, J. Spanier, \emph{The fractional calculus}, Academic Press, New York (1974); ISBN 978-0486450018.

\bibitem{OrsingherPolito2011} E. Orsingher, F. Polito, The Space-Fractional Poisson Process.
\emph{Statistics $\&$ Probability Letters} \textbf{82}, No 4 (2012), 852--858.

\bibitem{Podlubny1999} I. Podlubny, \emph{Fractional Differential Equations}, Academic Press, San Diego (1998); ISBN: 9780125588409.

\bibitem{PolitoScalas2016} F. Polito, E. Scalas, A generalization of the space-fractional Poisson process and its connection to some L\'evy processes. 
\emph{Electron. Commun. Probab.} \textbf{21}, No 20 (2016), 1--14.

\bibitem{PolitoTomovski2016} F. Polito, Z. Tomovski, Some properties of Prabhakar-type fractional calculus operators. \emph{Fract. Differ. Calc.} \textbf{6}, No 1 (2016), 73-94.

\bibitem{Prabhakar1971} T.R. Prabhakar, A singular integral equation with a generalized Mittag-Leffler function in the kernel. \emph{Yokohama Math. J.} \textbf{19} (1971), 7--15.

\bibitem{RepinSaichev2000} O.N. Repin, A.I. Saichev, Fractional Poisson law. \emph{Radiophys. and Quantum Electronics} \textbf{43} (2000), 738--741.

\bibitem{SaichevZaslavski1997} A.I. Saichev, G.M. Zaslavsky, Fractional kinetic equations: solutions and applications. \emph{Chaos} \textbf{7} (1997) 753--764.

\bibitem{SamkoKilbasMarichev1993} S.G. Samko, A.A. Kilbas, G.I. Marichev, \emph{Fractional Integrals and Derivatives, Theory and Applications}, Gordon and Breach, New York (1993); ISBN: 2881248640.

\bibitem{Sandev2017} T. Sandev, Generalized Langevin Equation and the
Prabhakar Derivative. \emph{Mathematics} \textbf{5}, No 4 (2017), Art. \# 66.

\bibitem{ScherLax1973} H. Scher, M. Lax, Stochastic Transport in a 
Disordered Solid. I: Theory.
\emph{Phys. Rev. B} \textbf{7}, No 10 (1973), 4491--4502.

\bibitem{ScherMontroll1975} H. Scher, E. Montroll, Anomalous transit-time dispersion in amorphous solids. \emph{Phys. Rev. B} \textbf{12}, No 6 (1975), 2455--2477.

\bibitem{Shlesinger2017} M. Shlesinger, Origins and applications of the Montroll-Weiss continuous time random walk. \emph{Eur. Phys. J. B} \textbf{90}, No 93 (2017), 1--5.

\bibitem{ShulkaPrajabati2007} A.K. Shukla, J.C. Prajapati, On a generalization of Mittag-Leffler function and its properties. 
\emph{J. Math. Anal. Appl.} \textbf{336} (2007), 797--811.

\bibitem{SolomonSwinner1993}  T. H. Solomon, E. R. Weeks, H. L. Swinney,
Observation of anomalous diffusion and L\'evy flights in a two-dimensional rotating flow.
\emph{Phys. Rev. Lett.} \textbf{71}, No 24 (1993), 3975--3978.

\bibitem{SungBarkaiSilbey2002}  J. Sung, E. Barkai, R. Silbey, S. Lee,
Fractional dynamics approach to diffusion-assisted reactions in disordered media.
\emph{J. Chem. Phys.} \textbf{116}, No 6 (2002), 2338--2341.

\bibitem{RiascosMateos2017} A. P. Riascos, J. L. Mateos, 
Emergence of encounter networks due to human mobility. \emph{PLoS ONE} 
\textbf{12}, No 10 (2017), Art. \# e0184532.

\bibitem{RiascosMateos2015} A. P. Riascos, J. L. Mateos, 
Fractional diffusion on circulant networks: emergence of a dynamical small world. 
\emph{J. Stat. Mech.} \textbf{2015}, No 7 (2015), Art. \# P07015.

\bibitem{RiascosMichel2018} A.P. Riascos, T.M. Michelitsch, B.A. Collet. , A.F. Nowakowski, F.C.G.A. Nicolleau, Random walks with long-range steps generated by functions
of Laplacian matrices. \emph{J. Stat. Mech.} \textbf{2018}, No 4 (2018), Art. \# 043404.

\bibitem{Zaslavsky2002}
G.M. Zaslavsky, Chaos, fractional kinetics, and anomalous transport. 
\emph{Phys. Rep.} \textbf{371} (2002), 461--580.


\end{thebibliography}
\end{document}